\documentclass[12pt]{iopart}
\usepackage{xcolor}
\usepackage{graphicx}
\usepackage{dcolumn}
\usepackage{bm}
\usepackage{multirow}
\usepackage[numbers,sort&compress,square]{natbib}
\usepackage{hyperref}
\usepackage[mathlines,running]{lineno}
\usepackage{subcaption}
\usepackage{tabularx}
\usepackage{verbatim}
\usepackage[mathlines,running]{lineno}

\begin{document}

\title[]{QoQ: a Q-transform based test for Gravitational Wave transient events}

\newcommand{\MIT}{LIGO Laboratory, Massachusetts Institute of Technology, Cambridge, MA 02139, USA}
\newcommand{\Portsmouth}{University of Portsmouth, Portsmouth, PO1 3FX, United Kingdom}
\newcommand{\Minnesota}{School of Physics and Astronomy, University of Minnesota,
Minneapolis, Minnesota 55455, USA}
\newcommand{\Montclair}{Montclair State University, 1 Normal Ave, Montclair, NJ 07043}
\newcommand{\Caltech}{LIGO Laboratory, California Institute of Technology, Pasadena, CA 91125, USA}
\newcommand{\UWM}{University of Wisconcin--Milwaukee, Milwaukee, WI 53201, USA}
\newcommand{\CITA}{Canadian Institute for Theoretical Astrophysics, Toronto, ON M5S 3H8, Canada}
\newcommand{\UofTPhysics}{Department of Physics, University of Toronto, Toronto, ON M5S 1A7, Canada}
\newcommand{\UofTAstro}{David A. Dunlap Department of Astronomy, University of Toronto, ON M5S 3H4, Canada}

\author{
Siddharth Soni$^{1}$,
Ethan Marx$^{1}$, 
Erik Katsavounidis$^{1}$,
Reed Essick$^{2,3,4}$
G. S. Cabourn~Davies$^5$, 
Patrick Brockill$^6$,
Michael W. Coughlin$^{7}$,
Shaon Ghosh$^{8}$,
Patrick Godwin$^9$
}
\address{$^1$\MIT}
\address{$^2$\CITA}
\address{$^3$\UofTPhysics}
\address{$^4$\UofTAstro}
\address{$^5$\Portsmouth}
\address{$^6$\UWM}
\address{$^7$\Minnesota}
\address{$^8$\Montclair}
\address{$^9$\Caltech}

\ead{siddsoni@mit.edu}
\vspace{10pt}
\begin{indented}
\item[]
\end{indented}

\begin{abstract}
The 
observation of transient gravitational waves is hindered by the presence of transient noise, colloquially referred to as glitches.
These glitches can often be misidentified as gravitational waves by searches for unmodeled transients using the excess-power type of methods and sometimes even excite template waveforms for compact binary coalescences while using matched filter techniques. They thus create a significant background in the searches.
This background is more critical in getting identified promptly and efficiently within the context of real-time searches for gravitational-wave transients.
Such searches are the ones that have enabled multi-messenger astrophysics with the start of the Advanced LIGO and Advanced Virgo data taking in 2015 and they will continue to enable the field for further discoveries.
With this work we propose and demonstrate the use of a signal-based test that quantifies the fidelity of the time-frequency decomposition of the putative signal based on first principles on how astrophysical transients are expected to be registered in the detectors and empirically measuring the instrumental noise.
It is based on the Q-transform and a measure of the occupancy of the corresponding time-frequency pixels over select time-frequency volumes; we call it ``QoQ''.
Our method shows a 40\% reduction in the number of retraction of public alerts that were issued by the LIGO-Virgo-KAGRA collaborations during the third observing run with negligible loss in sensitivity.
Receiver Operator Characteristic measurements suggest the method can be used in online and offline searches for transients, reducing their background significantly.

%
\end{abstract}

%
%
%

\section{Introduction and motivation}\label{sec:level1}
The direct detection of gravitational waves by ground-based interferometric detectors, Advanced LIGO (aLIGO)~\cite{TheLIGOScientific:2014jea} and Advanced Virgo (AdV) ~\cite{TheVirgo:2014hva} has burgeoned the field of gravitational-wave physics~\cite{Abbott:2016blz}.
Together with that, the first detection of a binary neutron star coalescence by LIGO-Virgo in association with a Gamma-Ray Burst~\cite{LIGOScientific:2017vwq} has also inaugurated a new era in multi-messenger observations enabled by gravitational-wave observations.
Three observing runs of the international network of gravitational-wave detectors have already taken place starting in 2015 and until 2020.
They are referred to as ``O1'', ``O2'' and ``O3''; they resulted in about 100 gravitational-wave candidates following thorough analyses of data conducted over these years~\cite{LIGOScientific:2016dsl, LIGOScientific:2018mvr, LIGOScientific:2021usb, LIGOScientific:2021djp, Nitz_2019, Nitz_2020, PhysRevD.100.023011, PhysRevD.101.083030, PhysRevD.106.043009}.
These gravitational-wave candidates are the result of collisions between compact binary objects such as black holes and neutron stars.
Additionally, during these observing runs, real-time searches were also conducted by the collaborations to disseminate transient event detection candidates to the broader astronomy community in order to enable their multi-messenger observations~\cite{LIGOScientific:2017ync}.
During the O3 run, 80 public alerts for gravitational-wave candidates have been distributed via the Gamma-Ray Burst Coordination Network (GCN).
Out of these, 24 were retracted on timescales of hours to days following their initial release, primarily reflecting manual examination of the events by the LIGO-Virgo Collaboration and deeming them to be of non-astrophysical origin.
Quantifying and automating the retraction process for this kind of event in upcoming observing runs of the LIGO-Virgo-KAGRA network of detectors has been the primary motivation and goal of this work.

The output of ground-based gravitational-wave detectors is strain amplitude, also referred to as $h(t)$.
This captures with the highest sensitivity 
the differential change in length of the kilometer-long arms of the interferometer.
The $h(t)$ is impacted by multiple sources of noise which can be broadly classified into two categories, short duration non-Gaussian transients, also known as $\textit{glitches}$, and noise that behaves in a Gaussian and stationary fashion over a given time and frequency range ~\cite{Davis:2021ecd, LIGOScientific:2019hgc, VIRGO:2012oxz, KAGRA:2020agh}.
These transients can adversely impact gravitational-wave (GW) searches both at the detection confidence level as well as in their source parameter estimation, including localization.~\cite{Canton:2013joa,TheLIGOScientific:2017lwt,Pankow:2018qpo,Powell:2018csz,Chatziioannou:2021ezd,Payne:2022spz,Macas:2022afm, LIGOScientific:2016gtq}.
Aside from their impact to searches, identification of transient noise may lead to studies on their coupling mechanism in the detector and subsequent hardware changes that can help remove or at least reduce such noise sources~\cite{Soni:2020rbu}.

The task of identifying transient gravitational-wave candidate events utilizes the $h(t)$ time-series. It is performed by a variety of transient-finding search pipelines developed by the LIGO-Virgo-KAGRA collaborations, as well as the broader community.
In this paper, we focus on low-latency 
searches~\cite{DalCanton:2020vpm,Sachdev:2019vvd,
Aubin:2020goo,Hooper:2011rb}.
These searches are primarily responsible for producing public GCN alerts in close to real time in order to allow for their multi-messenger follow-up.
The LIGO-Virgo-KAGRA (LVK) collaboration also performs searches offline~\citep{Davies:2020tsx,Messick:2016aqy, Aubin:2020goo, Venumadhav:2019lyq} with refined versions of their online pipelines. 
Such refinement allows for, among other things, improved sensitivity, better noise rejection, although it often comes with higher computational complexity and slower turn-around times.
The broader GW community outside the LVK also analyses the interferometric data once they become publicly available, with redundant and complementary results obtained.
In the typical searches for compact binary coalescences, the waveform of such signals can be modelled using post-Newtonian and numerical relativity methods, thus resulting in a family of signals spanning a space parameterized by masses and spins of the binaries; this is referred to as \textit{template banks}.
The various search pipelines employ what is known as \textit{match filtering} to look forGW signals in the data, by matching the incoming strain with waveforms in such template banks \cite{Usman:2015kfa, Messick:2016aqy, Aubin:2020goo}.
However, the presence of noise transients in the data complicates this process as they may mask and sometimes mimic a true GW signal \cite{LIGOScientific:2017vwq}.
To remedy this, the search pipelines use a \textit{chi-squared test} to differentiate the time-frequency distribution of power in real GW signals and noise transients \cite{Allen:2004gu, Nitz:2017lco}.
Even with all these precautions, transient noise can trigger a pipeline alert for a GW candidate event, which consequently has to be retracted.
The decision to retract an event is carried out via human intervention with minimal quantitative analysis.


In this article, we present a tool to distinguish binary black hole (BBH) signals from transient noise based on the energy distribution in the signal across the time-frequency plane, as obtained via the Q-transform~\cite{Chatterji:2004qg}.
In Section \ref{retr_events_sec}, we discuss events retracted during O3. In Section \ref{gauss_noise_sec}, we look at the energy distribution of Gaussian and transient noise and describe in detail the Q-occupancy (QoQ) test. In Section \ref{pix_occ_sec}, we discuss its application on O3 low latency astrophysical candidates and retracted events. In Section \ref{mdc_pycbc_sec}, we discuss our QoQ test analysis on the Mock Data Challenge (MDC) Injection Data Set and PyCBC background events. In Section \ref{O3_sec}, we extend the analysis of Section \ref{pix_occ_sec} to events found in offline analysis in O3. Finally we conclude and summarize in Sec \ref{disc_sec}.

\section{O3 retracted events}\label{retr_events_sec}
During the third Observing run (O3), a total of 80 events were identified in low latency, of which 23 were later retracted as their origin was found to be environmental or instrumental rather than astrophysical.
An additional event was retracted as the alert was sent out due to an error~\footnote{https://gracedb.ligo.org/superevents/S190405ar/view/} \cite{LIGOScientific:2021djp,LIGOScientific:2021usb,dcc:CBC_in_O3a, dcc:CBC_in_O3b, LIGOScientific:2023vdi}. 
Offline re-analysis of the O3 data refined the admission criteria of astrophysical events, including the requirement of the probability of an event being astrophysical to be greater than $0.5$ and having a False Alarm Rate (FAR) of less than 2 per day; this led to the GWTC-3 catalog~\cite{LIGOScientific:2021usb}.
Of the non-retracted 56 events identified in low latency, $44$ events were also identified by the offline re-analysis and thus included in the GWTC-3 catalog~\cite{LIGOScientific:2021usb}.
A long-term solution to reduce the number of retractions is to identify the transient noise coupling in the detector and remedy it by making hardware changes \cite{Soni:2020rbu}. During O3b, after the reaction chain tracking was employed at LIGO Livingston Observatory (LLO) and LIGO Hanford Observatory (LHO) to diminish the slow scattering transient noise, the average fraction of retracted O3b public alerts before and after dropping from $0.55$ to $0.21$ ~\cite{LIGOScientific:2021djp}. However, reduction of transient noise is only sometimes possible as the noise originates in the complex instrumentation and environment of the detector, and new categories of transient noise may always show up as the sensitivity of the detector improves and new subsystems get added ~\cite{Cabero:2019orq, Soni:2021cjy,Glanzer:2022avx,glanzer2023noise}.  
So, a need for prompt, quantifiable, and automated ability to address noise transients and their impact on public alerts from the LIGO-Virgo-KAGRA network remains.
\begin{figure}[h]
\captionsetup[subfigure]{}
   \centering
    \begin{subfigure}{0.45\textwidth}
        \centering
         \includegraphics[width= 0.9\textwidth]{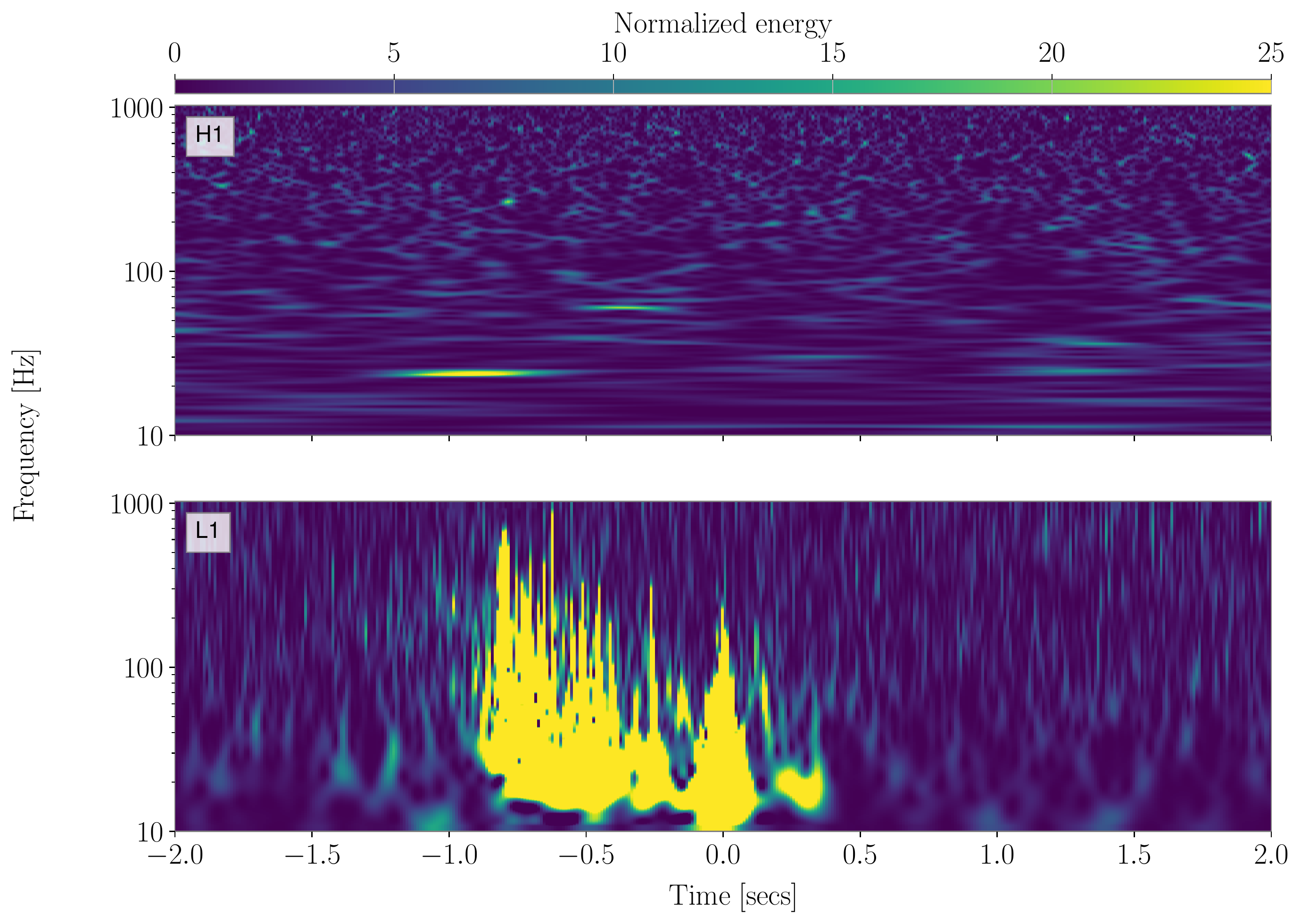}
         \caption{Retracted event S191117j \cite{GCN26254}}
         \label{fig:retr_1258}
    \end{subfigure}
    \hspace{2em}
    \begin{subfigure}{0.45\textwidth}
        \centering
         \includegraphics[width =0.9\textwidth]{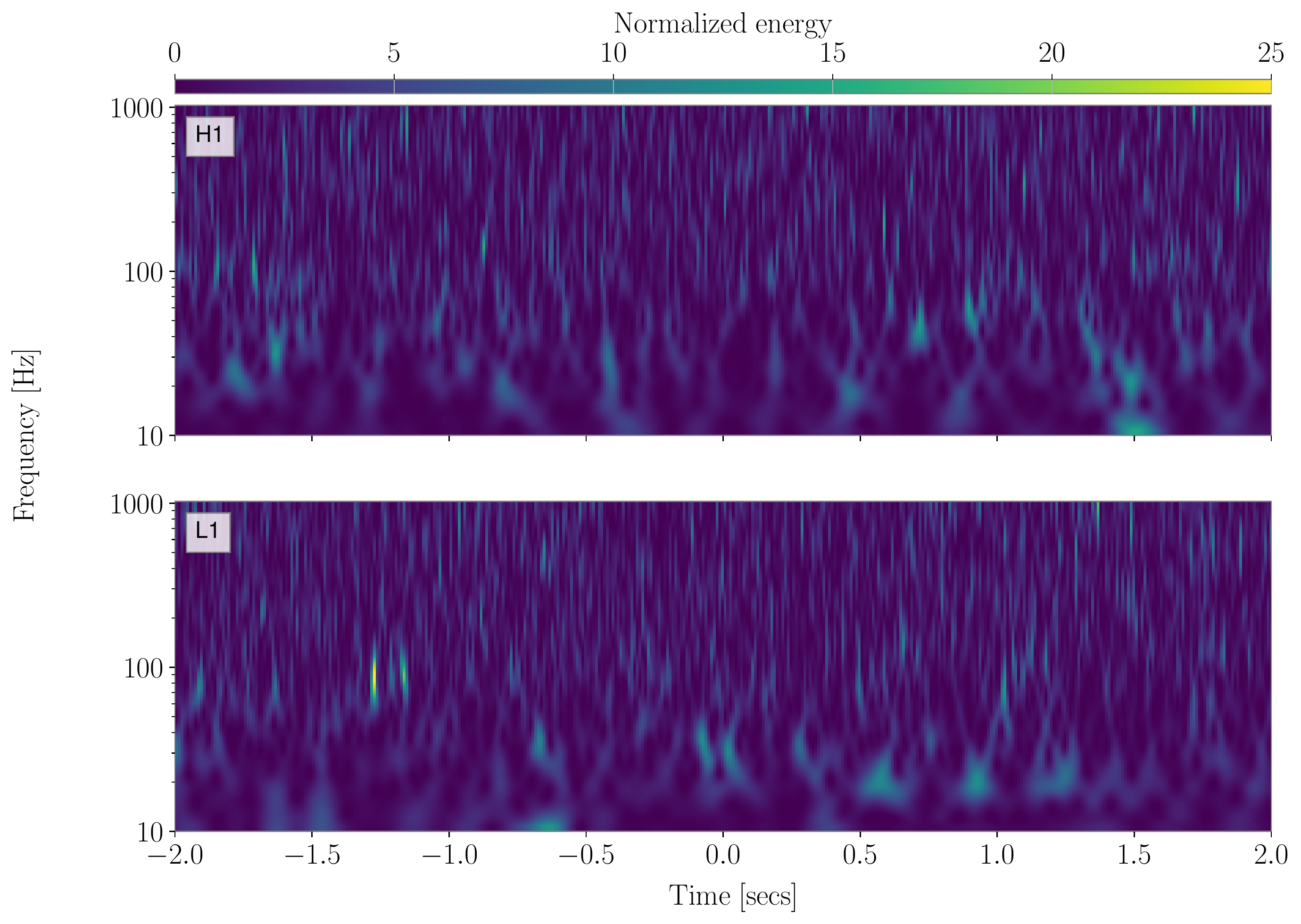}
         \caption{Retracted event S190808ae \cite{GCN25301}}
         \label{fig:retr_12493}
    \end{subfigure}
    \caption{Q-transform visualizations of example retracted events from O3. \emph{Left}: L1 data has a loud glitch at the time of event S191117j \cite{GCN26254}. \emph{Right}: The retracted event S190808ae do not contain any visible presence of transient noise.
    }
    \label{fig:retractions}
\end{figure}

Retracted events during O3 can be broadly classified into two categories: events with clear evidence of loud transient noise in one or more detectors and low SNR events with almost no visible presence of transient noise. Fig \ref{fig:retractions} shows an example from each category. About half of the O3 retractions belong to the first category. We aim to separate these false positives from astrophysical transient candidate events. The QoQ test explained in Section \ref{pix_occ_sec} later on in the paper is a method for distinguishing transient noise from astrophysical events using the Q-transform and the time-frequency pixel-energy distribution measure that it provides. There has been a growing interest among researchers in solving such problems \cite{Vazsonyi:2022jul, Alvarez-Lopez:2023dmv, Kapadia:2017fhb}.

\section{The Q-transform in gravitational wave data analysis}\label{gauss_noise_sec}

A commonly used analysis tool for identifying and characterizing transients both in the gravitational-wave strain channel and in the wealth of auxiliary channels invoked in the interferometry is the Q-transform~\cite{gabor,brown}.
It is a modification of the short Fourier transform that covers the time-frequency plane with pixels of constant $Q$ (quality factor) by invoking analysis windows with durations that vary inversely proportional to the frequency.
It can be shown that it is equivalent to a matched filter search for minimum uncertainty waveforms~\cite{Chatterji:2004qg,chatterji_2005}.
A search for gravitational-wave bursts using the Q-transform was first implemented in the early science runs of initial LIGO \cite{Chatterji:2004qg, chatterji_2005}.
The implementation of Q-transform as an unknown-morphology transient-finding method generally tiles the time-frequency plane using a range of quality factors $Q$.
We will refer to these tiles as {\it{pixels}}.
The squared magnitude of their (discrete) Q-transform coefficient is a measure of the signal energy, and we will refer to it as {\it{pixel energy}}.
For white noise, such pixel energy at a given frequency is exponentially distributed.
Transients are generally identifiable as outliers in the pixel energy distribution via invoking a threshold and some clustering algorithm in order to achieve a desired false alarm rate on the assumption of Gaussian (or otherwise directly measured from the data) noise.

\begin{figure}
    \centering
    \includegraphics[width=0.75\textwidth]{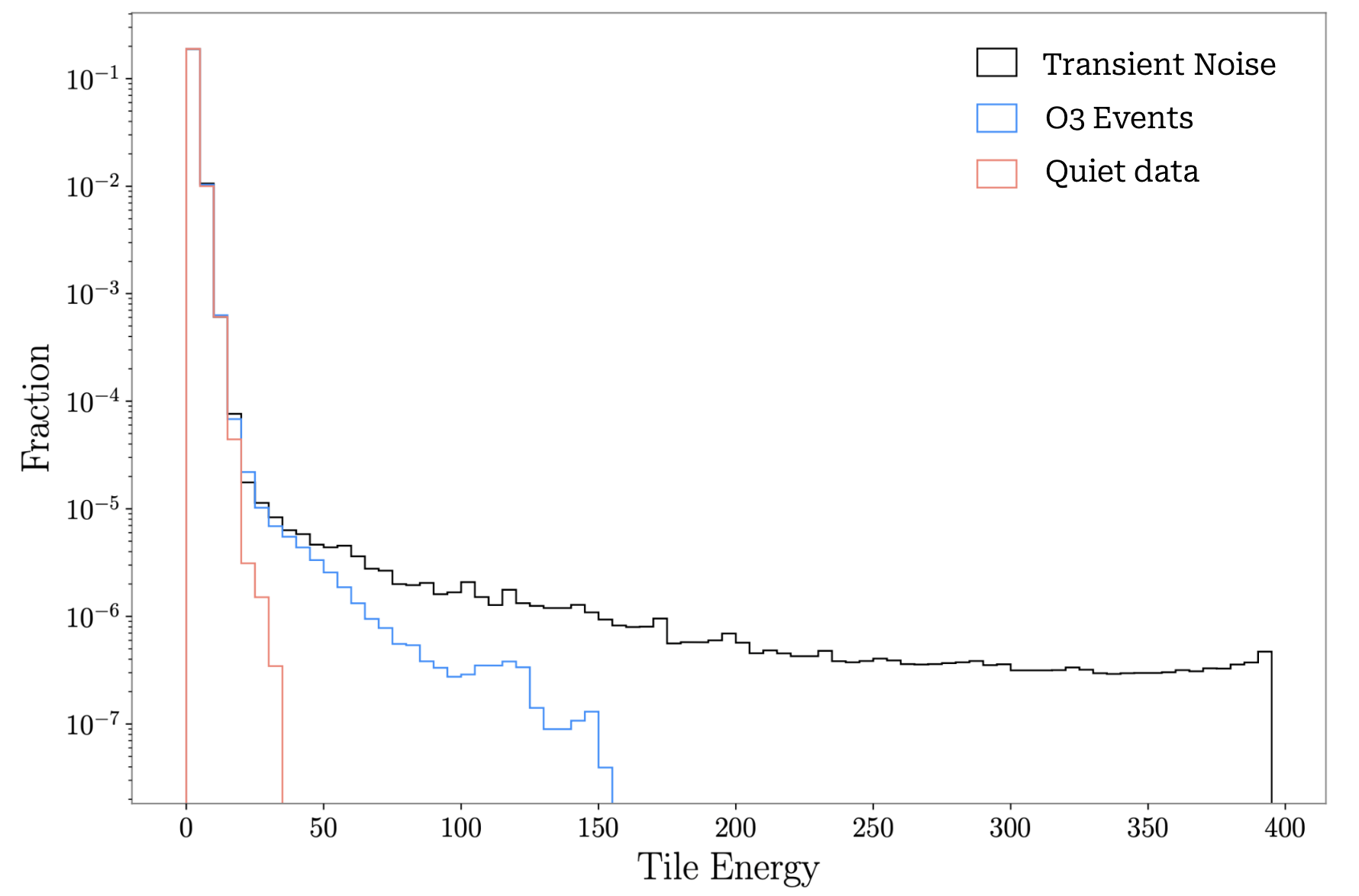}
    \caption{Q-transform pixel-energy distribution comparison of detector data around noisy, O3 catalog, and quiet times in the LIGO detectors. ``Noisy'' times are identified by {\it{Omicron}} as being subject to a glitch while ``quiet'' times are selected via randomly sampling times of the LIGO detectors data taking over which no glitch by {\it{Omicron}} was identified. As compared to quiet times, noise transients or short duration glitches appear in the tail of the pixel-energy distribution.
    }
    
    \label{fig:tile_energy_comp}
\end{figure}

This can be seen in Fig \ref{fig:tile_energy_comp} where we histogram the energy of all the pixels generated from a 4s long Q-transform of 16 ``noisy'' and 16 ``quiet'' times of the LIGO detectors.
The noisy times refer to ones identified by the {\it{Omicron}}~\cite{Robinet:2020lbf} method as having a glitch of signal-to-noise (SNR) ratio of at or above $6$ within such time-window, while ``quiet'' times have been randomly sampled from the LIGO detectors data taking corresponding to livetime over which no glitch has been identified by {\it{Omicron}}.

\subsection{Pixel occupancy}

Our proposed test for gravitational-wave transient events uses as a discriminant the fraction (in percentage) of pixels within a few time-frequency volumes that exceed pre-set values of pixel energy.
We refer to this quantity as ``occupancy'' or ``density'' of pixels, thus the name of our method as ``Q-occupancy'', or {\it{QoQ}}.
Astrophysical transients of known morphology, including compact binary coalescences (CBC), have analytically calculable occupancy values (given the astrophysical source parameters).
For those of unknown morphology, it is a priori impossible to discriminate them from non-Gaussian noise artifacts via a single {\it{QoQ}} measurement.
In this case, in order to utilize {\it{QoQ}}, some basic (or non) assumptions need to be made on the physics and the detection of the (unknown morphology) astrophysical signal one is after.  Astrophysical signals are generally louder than Gaussian noise but usually not as loud as the glitches responsible for the retracted events. As shown in Fig \ref{fig:tile_energy_comp}, the pixel-energy distribution of the true events extends beyond the Gaussian noise but not as far as the transient noise. And the kind of noise artifacts in gravitational-wave interferometers we are primarily gearing our method to identify has an extended structure in the time-frequency plane and often exceeds pixel energies expected from astrophysical populations we currently know \cite{Davis:2021ecd}.
This is learned by the noise data the instruments record and its implications to searches is assessed via standard confusion matrix and Receiver Operating Characteristic (ROC) analysis under some assumption on the astrophysical signal in consideration.

In this work we utilized the {\it{GWpy}} \cite{gwpy} and {\it{Omicron}} \cite{Robinet:2015om,Robinet:2020lbf} implementation of the Q-transform.
We have used a range of quality factors $Q$ from $4$ to $64$ and frequencies from $10$ Hz to $1024$ Hz to logarithmically tile the time-frequency plane.
Analysis of the {\it{h(t)}} time series via the Q-transform results to ``triggers'' corresponding to transients present in the data.
Such triggers are described with a few parameters such as the central event time, central frequency, quality factor {\it{Q}}, and signal-to-noise ratio; we will use these for the main feature description of the transients we will be analyzing.
Additionally, the full time-frequency decomposition with all pixel energy values across resolutions is made available and is what we use for deriving the {\it{QoQ}} quantities.

Several options exist on how to define the time-frequency volume over which to measure {\it{QoQ}}.
We first need to pick a threshold on the pixel-energy at or above which we will consider the pixels as ``loud'' and thus counted toward the {\it{QoQ}}.
We generally need to pick values above the Gaussian noise level, although the exact threshold has to be informed by the character of the noise and the corresponding impact on the misclassification of signals from astrophysical populations. 
The overall time-frequency volume over which $h(t)$ is analyzed can vary significantly depending on the astrophysical search within which noise rejection via this method is being implemented.
In this analysis, we will focus on transient gravitational waves from binary black hole (BBH) mergers typically lasting from tens of milliseconds to a few seconds.
Extension of the method to other astrophysical populations is straightforward.

\begin{figure}[h]
    \centering
    \includegraphics[width=0.85\textwidth]{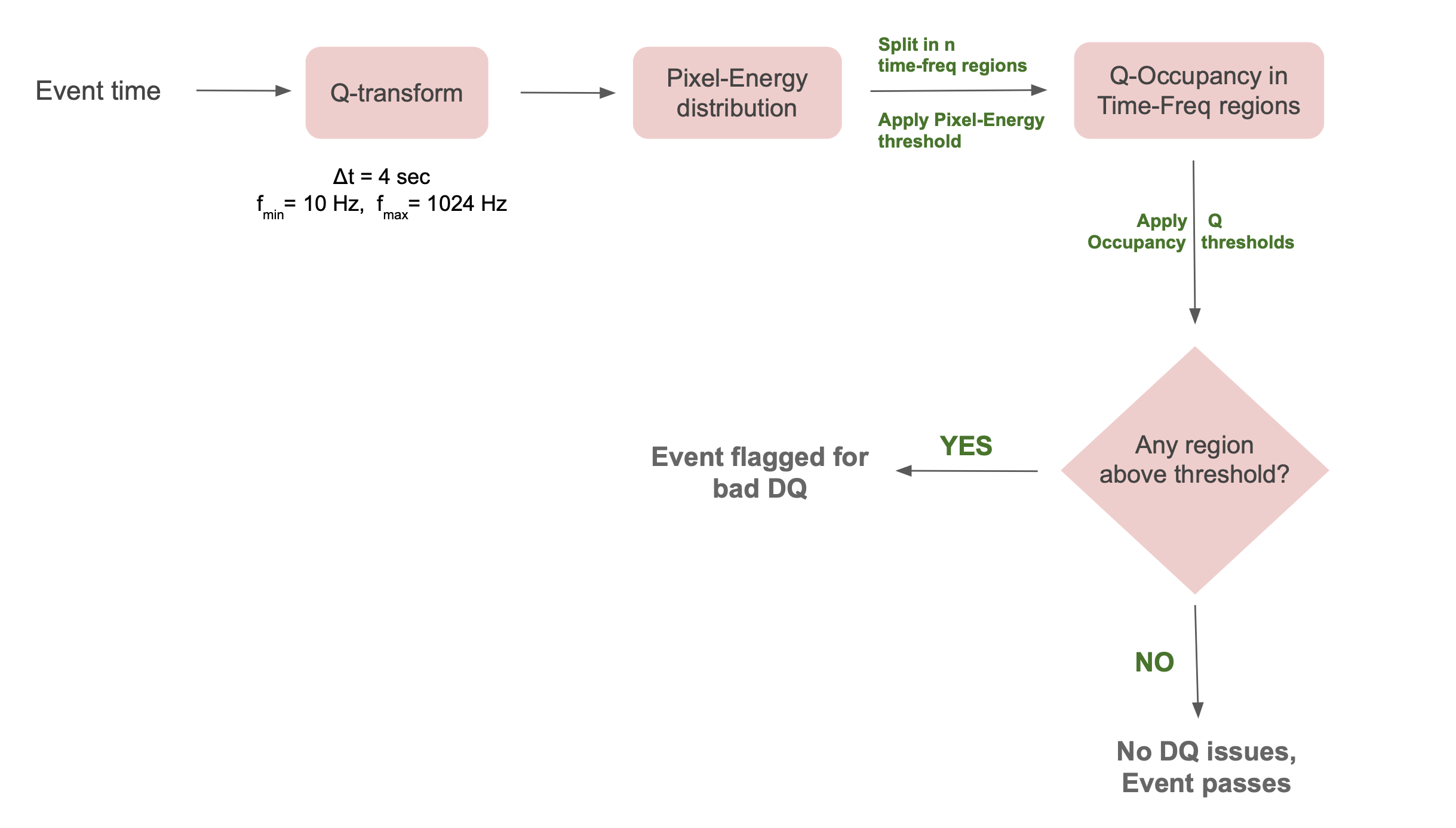}
    \caption{Flowchart of the QoQ test. Given an event time, the tool makes a 4 second long Q-transform. We then split the pixel-energy of this Q-transform into multiple time-frequency regions and apply a pixel-energy threshold. This gives us, for each time-freq region, fraction of pixels above the pixel-energy threshold i.e. its Q-occupancy. The next step is to apply a QoQ threshold on each of these regions. If any of these time-freq regions are above its QoQ threshold, the event is flagged for bad data quality, otherwise it passes the QoQ test.}
    \label{fig:flow_chart_process}
\end{figure}

We use a duration of 4 seconds set by the scale of BBH events for the Q-transform and restrict the analysis to the frequency band of $10$--$1024~\mathrm{Hz}$.
The frequency span of the search is motivated by the overall frequency content of noise artifacts in aLIGO and AdV as well as that of the BBH and CBC -- in general -- signals.
This Q-transform is then sub-divided into $n$ (disconnected, i.e., mutually exclusive) time-frequency regions spanning the original time-frequency volume. For each of these $n$ regions, the fraction (in percentage) of pixels above a given pixel-energy threshold is computed.
We will refer to this fraction as the {\it{QoQ}} value.
If the {\it{QoQ}} exceeds a certain threshold in a given region, we use this to infer that the corresponding region is contaminated with noise. This procedure is shown as a flow chart in Fig \ref{fig:flow_chart_process}. 

The method allows for a different {\it{QoQ}} threshold to be used in each of the $n$ sub-divisions of the time-frequency volume one starts with.
In our approach, currently, any one of such $n$ sub-divisions exceeding the {\it{QoQ}} threshold flags the event for further inspection.
Clearly, more complicated logic can be implemented, making certain assumptions for the hypothetical astrophysical signal as well as for the noise behavior that may be targeted via such an approach.
This procedure is shown as a flow chart in Fig \ref{fig:flow_chart_process}.
There is an additional conceptually different approach in the sub-division of the original time-frequency volume, namely in $m$ overlapping rather than disconnected regions.
We invoke this approach as well since, as we will show below, it offers a unique handle in identifying astrophysical events that may be overlapping with noise artifacts.
\begin{figure}[h]
    \centering
    \includegraphics[width=0.85\textwidth]{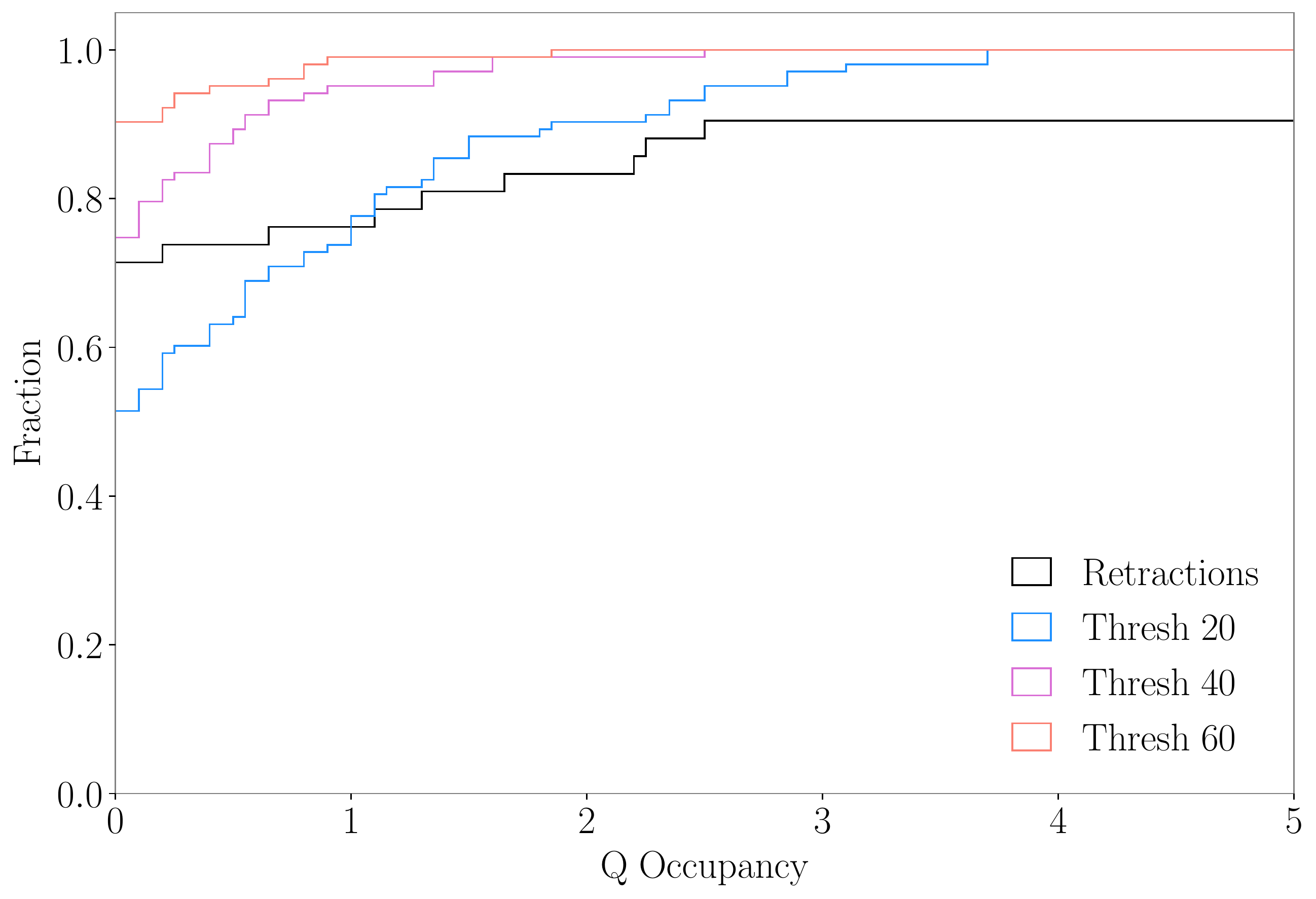}
    \caption{ Q occupancy distribution of 1 second window surrounding the event time of O3 online public alerts and O3 retracted events, for different values of pixel-energy thresholds. The retracted events with clear presence of transient noise have much higher Q occupancy values compared to the astrophysical candidates. For the O3 online candidates, the max Q occupancy is below 4, for the retractions the distribution extends to Q occupancy of much higher than 5 shown here. }
    \label{fig:Q-occupancy_dist}
\end{figure}

To decide the values of pixel-energy and QoQ thresholds to be used in the QoQ test, we plot the cumulative distribution of QoQ values in the 1-second window surrounding the event time of O3 low latency candidates and O3 retracted events.
Fig \ref{fig:Q-occupancy_dist} shows this QoQ distribution of O3 low latency candidates for different pixel-energy thresholds and of O3 retracted events for a pixel-energy threshold of 60. As we can see from this figure, in the case of O3 online candidates, for a pixel-energy value of 60, the maximum QoQ value for the central 1-sec window is 2.0, we use these values as our thresholds for the pixel-energy and QoQ, respectively for the QoQ analysis on O3 online candidates next.

\section{O3 online candidates analysis}\label{pix_occ_sec}
We apply the QoQ test on the O3 online candidates. This set includes 56 non retracted and 23 retracted events. We first perform the overlapping and non overlapping analysis individually in \ref{overlap_analysis} and \ref{sec:nonoverlap} respectively and then combine the two analyses to obtain the results shown in Tab \ref{tab_retrs_O3}.

\begin{figure}[h]
    \centering
    \begin{minipage}[t]{.5\textwidth}
        \centering
        \includegraphics[width=0.9\linewidth]{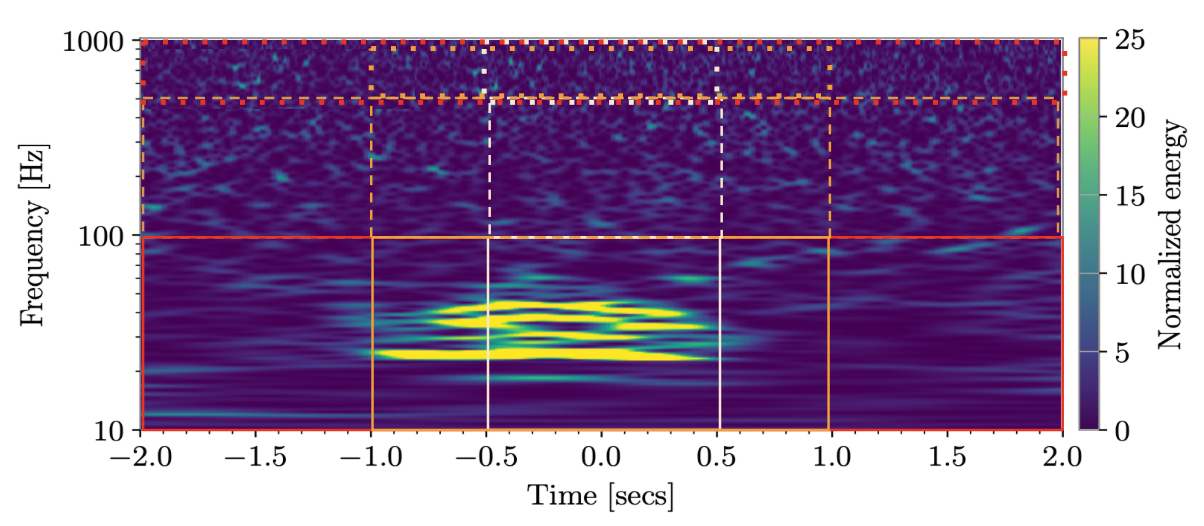}
    \end{minipage}%
    \begin{minipage}[t]{.5\textwidth}
        \centering
         \includegraphics[width=0.9\linewidth, height=0.4\linewidth]{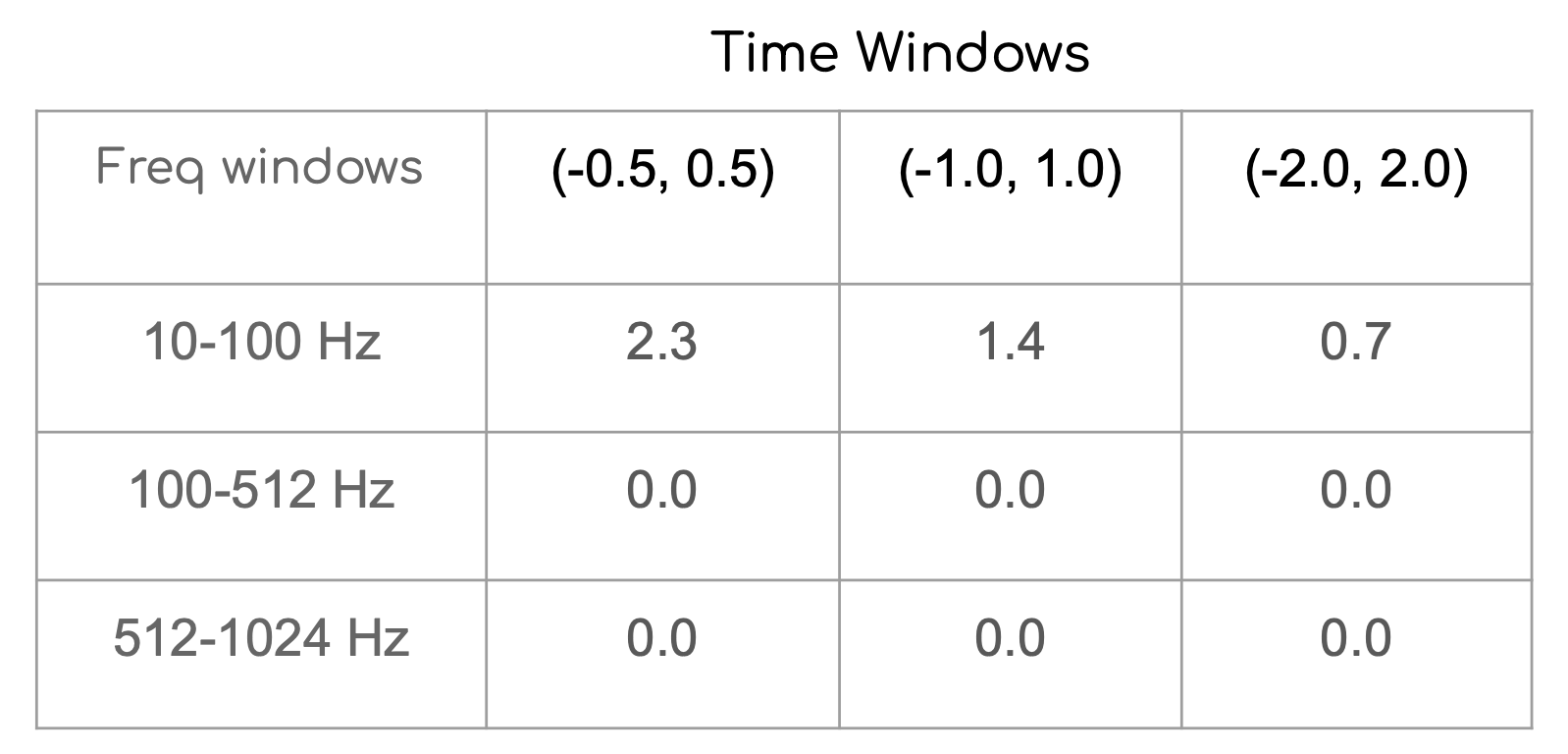}
    \end{minipage}
    \caption{\emph{Left}: A Q-transform divided into 9 time frequency blocks. These blocks which are 1, 2 and 4 seconds long are centered at $t=0$ and overlap in time. They cover three frequency bands from $\mathrm{10-100}$ Hz, $\mathrm{100-512}$ Hz and $\mathrm{512-1024}$ Hz.
    \emph{Right}: This table shows the fraction of pixels with energy above the pixel-energy threshold of 60 in each of the 9 time-frequency regions.}
    \label{fig:time_freq_regiona}
\end{figure}

\subsection{Overlapping time windows and O3 online candidates analysis}\label{overlap_analysis}
The frequency axis in the Q-transform shown on the left in Fig \ref{fig:time_freq_regiona} is divided into three bands, ($\mathrm{10-100}$ Hz), ($\mathrm{100-512}$ Hz), and ($\mathrm{512-1024}$ Hz) and the time axis into three windows (-0.5 s, 0.5 s), (-1.0 s, 1.0 s) and (-2.0 s, 2.0 s). The intersection of these time and frequency windows gives us a total of 9 regions. For each region, we calculate the fraction of pixels in the given region with a pixel-energy value above the pixel-energy threshold. The fractions for each region are shown on the right in Fig \ref{fig:time_freq_regiona} for a pixel-energy threshold of 60. For example, the value $2.3$ in the first row and first column indicates that for the 1-second long window in the frequency band, $10$--$100~\mathrm{Hz}$, $2.3$ percent of the pixels have pixel-energy above 60. The QoQ values for GW signals are smaller in comparison, and a suitable threshold can be used to separate the transient noise-generated false alerts from true GW candidate events.

\begin{figure}[h]
\captionsetup[subfigure]{}
  \centering
    \begin{subfigure}{0.45\textwidth}
        \centering
         \includegraphics[width =0.9\textwidth]{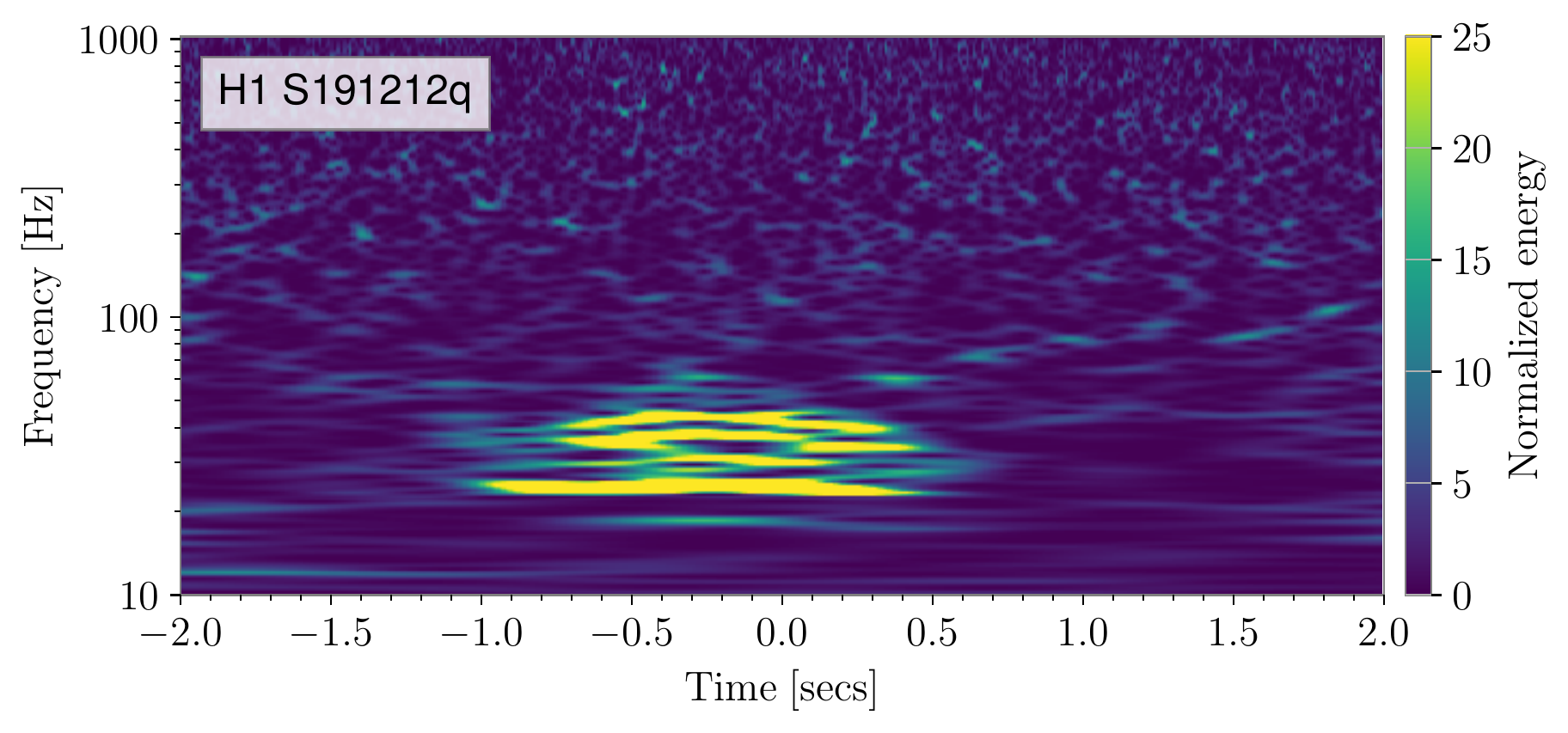}
         \caption{S191212q \cite{GCN26394}}
         \label{fig:retr_12601}

    \end{subfigure}
    \hspace{1em}
      \begin{subfigure}{0.45\textwidth}
        \centering
         \includegraphics[width =0.9\textwidth]{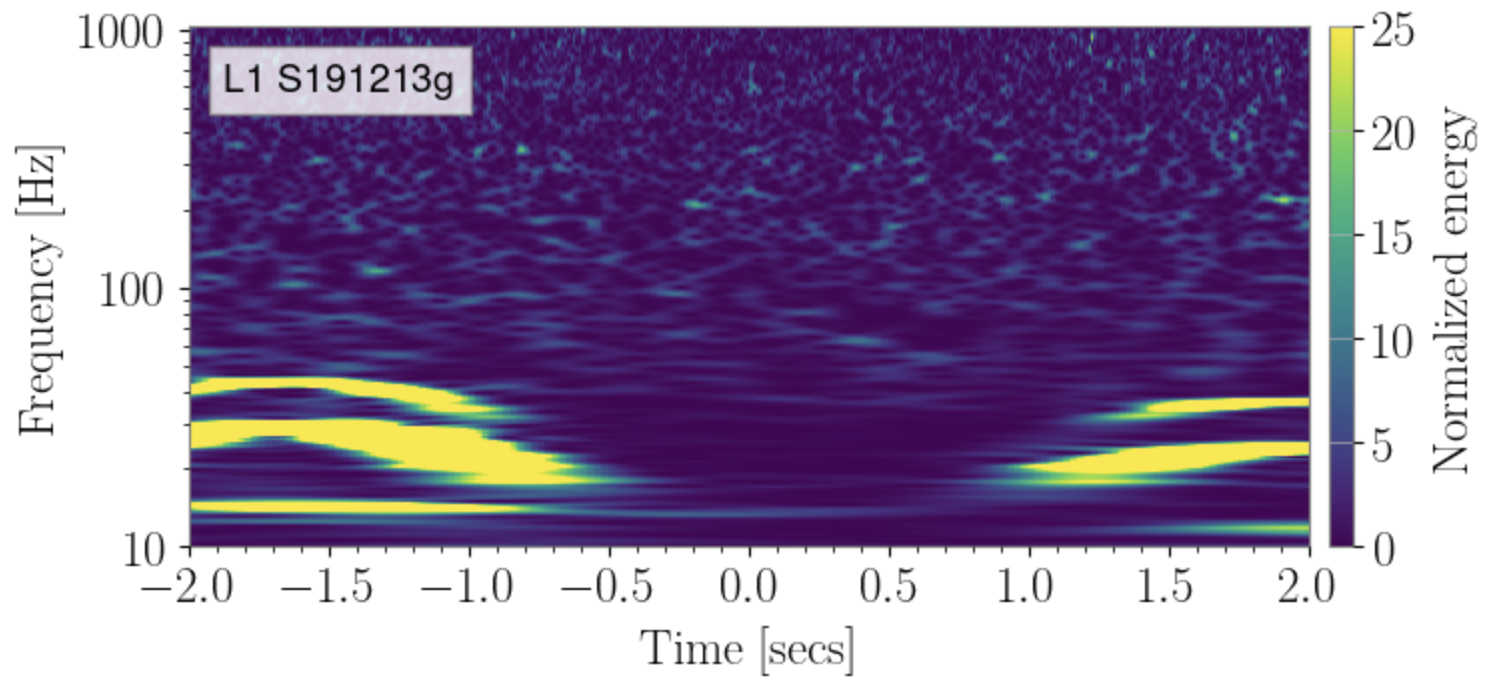}
         \caption{S191213g \cite{LIGOScientific:2021djp,GCN26399}}
         \label{fig:online_191213g}

    \end{subfigure}
    \caption{\emph{Left}: The H1 data quality around the time of retracted event S191212q \cite{GCN26394} is impaired by the presence of Scattering glitch. The transient noise is clearly visible in the Q-transform of H1. \emph{Right}: The O3 online event S191213g was found with FAR of 1.1 per year in low latency and a public alert was issued. The surrounding data quality is heavily contaminated with scattered light glitches. The S3 offline analysis did not identify this event as a significant candidate \cite{LIGOScientific:2021djp}.}
    \label{fig:retractions_flagged_notflagged}
\end{figure}

We use the \textbf{QoQ thresholds of 2, 1, and 1 for the three time windows $t_{1}=1$ sec, $t_{2}=2$ sec, and $t_{3}=4$ sec respectively}. To reduce the chances of flagging a true GW signal, the threshold on the central 1-sec window is higher as the majority of BBH signals fall within this. With these QoQ thresholds and a pixel-energy threshold of 60, 9 out of the 23 retractions and 1 of the 56 O3 online transient astrophysical candidates are flagged. \ref{appen_a} contains more details for each of the flagged events by the QoQ analysis at this stage
Fig \ref{fig:retractions_flagged_notflagged} shows two O3 online events that our method classifies as noise. More than half of the retracted events do not contain visible glitch activity, and the QoQ test does not flag these events. Out of 23 retractions, 14 were not flagged as noise by the QoQ test as they are low SNR events at one or both detectors.

\subsection{Non overlapping time windows and O3 online candidates analysis}\label{sec:nonoverlap}
\begin{figure}[h]
    \centering
    \begin{minipage}[t]{.5\textwidth}
        \centering
        \includegraphics[width=0.9\linewidth]{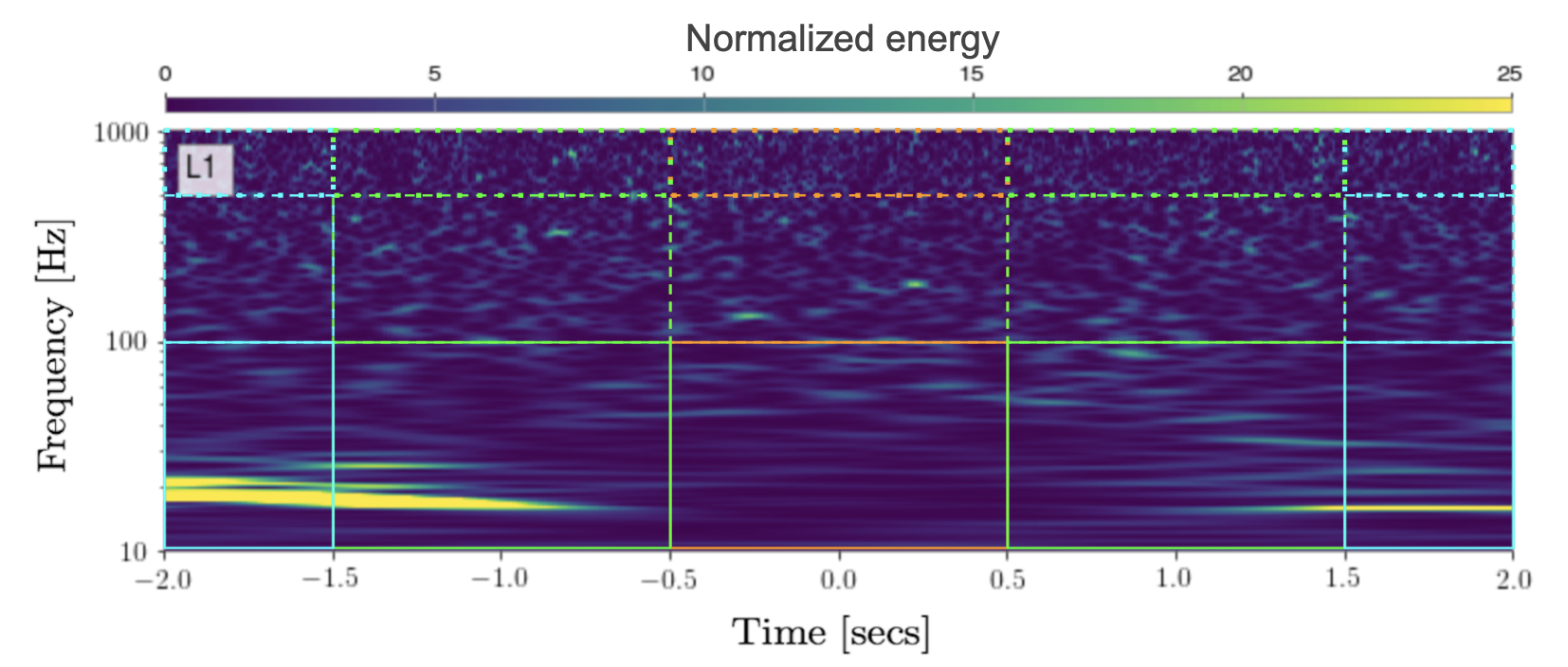}
        \label{fig:sub1}
    \end{minipage}%
    \begin{minipage}[t]{.5\textwidth}
        \centering
         \includegraphics[width=0.98\linewidth, height=0.38\linewidth]{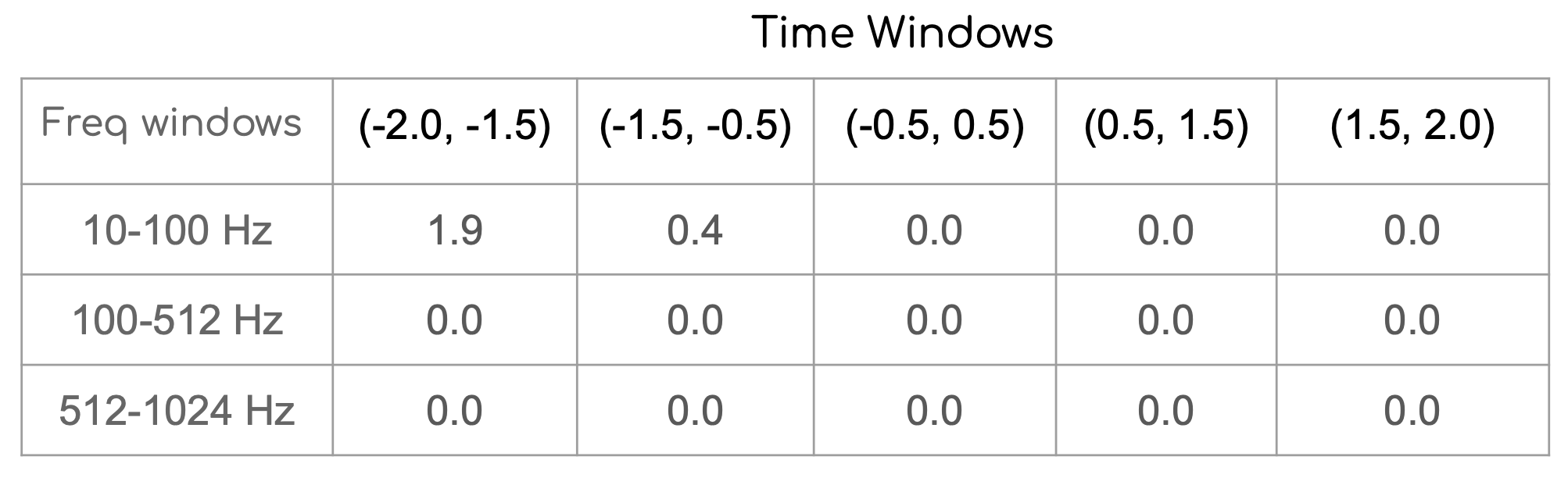}
       
        \label{fig:sub2}
    \end{minipage}
     \caption{\emph{Left}: A Q-transform divided into 15 time frequency regions. There is no overlap in time among these blocks. The first time window is half a second long and extends from -2.0 to -1.5 sec, the next three windows are each 1 sec long and the final region is again 0.5 secs long and extends from 1.5 sec to 2.0 sec. The division along the frequency axes is similar to the overlapping case. \emph{Right}: The table shows the fraction of pixels with energy above the pixel-energy threshold of 60, for each of these regions. There is some transient activity in the first one and a half second as seen in the Q-transform on left, this is reflected in the table on right.}
     \label{fig:time_freq_regionb}
\end{figure}

Another way to create the time-frequency regions in the Q-transform is with windows that do not overlap in time. Compared to the overlapping analysis, there is no change in frequency delineation, but the time axis is now split into five windows: (-2.0 s, -1.5 s), (-1.5 s, -0.5 s), (-0.5 s, 0.5 s), (0.5 s, 1.5 s) and (1.5 s, 2.0 s). In this case, we get a total of 15 time-frequency regions. And again, we calculate the fraction of pixels with a pixel-energy value above a certain pixel-energy threshold for each region. For example, fig \ref{fig:time_freq_regionb} shows the 15 time-frequency regions with no overlap in time and the QoQ of pixels for each region using a pixel-energy threshold of 60. \par
         

In many cases, environmental or instrumental noise artifacts adversely impact the data quality a few seconds before or after a GW chirp signal. These events may not be retracted, but they require closer inspection for any noise coupling with the auxiliary channels. The gravitational waveforms resulting from the merger of binary black holes are of the order of a few tens of milliseconds to a few seconds in the sensitive frequency band of ground-based detectors. This means any signal energy outside this time window can be attributed to a non-astrophysical phenomenon. The non-overlapping time windows method can be conveniently used to catch such events. Any events for which the QoQ in the central time window (-0.5, 0.5) is above 2 (same as the overlapping case) or in any of the non-central windows is above 0 can be flagged for stricter event validation. In this qualitative way, the non-overlapping time windows method differs from the overlapping time windows. 
 
The QoQ thresholds used in this case are $[0,0,2,0,0]$, i.e. events with QoQ \textbf{above $2\%$ for the central window ($t3$) or any non-zero value in any of the non-central windows will be flagged as transient noise}.

The O3 retracted events flagged by this method are the same as those flagged by the overlapping time windows method shown in \ref{overlap_analysis}. However, unlike the overlapping case, this method finds the presence of transient noise in three GW candidates found in low latency in O3. Unsurprisingly, none of the central time window (t3) values are flagged since this window is equivalent to the central window (t1) used in the overlapping method. However, the non-central windows flag a total of three O3 online astrophysical candidates. All of these three events are real GW signals polluted by the presence of noise artifacts in their vicinity at one or both detectors. \ref{appen_a} contains more details on these events.
During O3, these events were examined for any potential correlation with the auxiliary channels and underwent a careful process of offline event validation \cite{Abbott:2020niy,Davis:2022ird}. To summarize, the non-overlapping time windows method effectively catches both false pipeline alerts due to transient noise as well as real astrophysical signals polluted by nearby transient noise. 



\begin{figure}[h]
    \centering
    \includegraphics[width=0.75\textwidth, height = 0.22\textwidth]{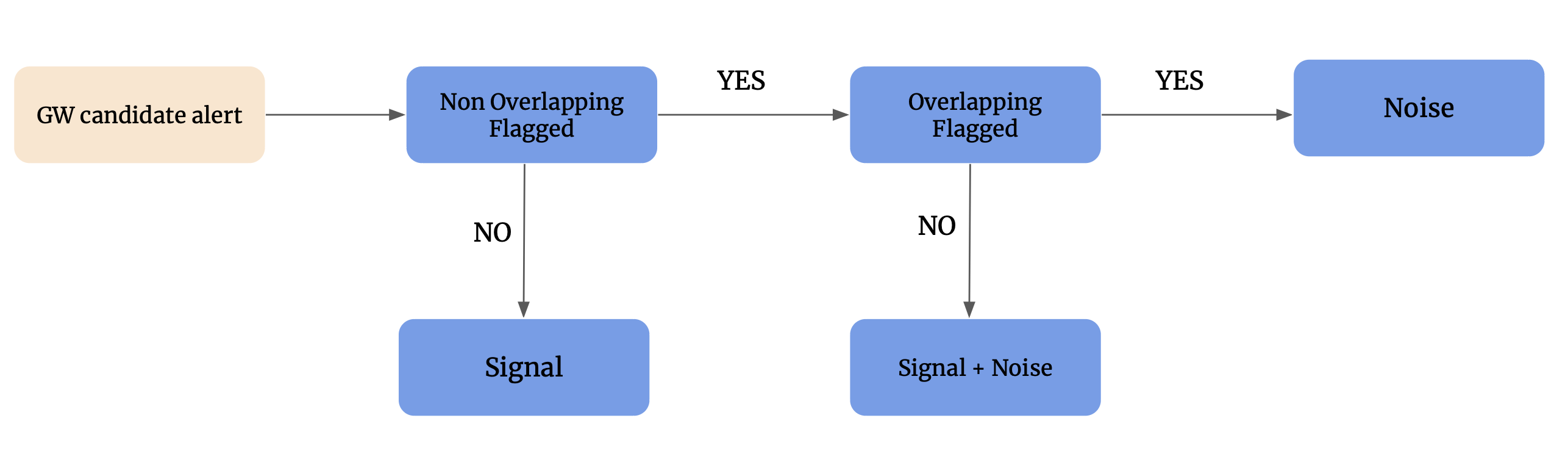}
    \caption{To evaluate the data quality of a GW candidate event, we first feed the event to Non Overlap QoQ test, if the event is not flagged at this stage, it is considered a ``Signal''. However if the event is flagged, it is then analyzed by the Overlap QoQ test. If the event is flagged at this stage, it is considered a ``Noise'', otherwise it is classified as a ``Signal with presence of transient noise''. }
    \label{fig:flowchart}
\end{figure}

Based on this analysis, we can use binary decision-making as shown in Fig \ref{fig:flowchart} to predict the nature of a GW candidate alert.  
Suppose an event is flagged by both the overlapping and the non-overlapping time windows analyses. In that case, such an event is likely a noise transient, and caution should be exercised before disseminating the information to the public. If the event is only flagged by the non-overlapping time windows method, then it is likely an astrophysical signal along with some transient noise nearby. Conversely, if an event is not flagged by the non-overlapping method, then it is an astrophysical event. 
Table \ref{tab_retrs_O3} shows the results of QoQ test on O3 events found in low latency and O3 retractions at different stages of the flowchart in Fig \ref{fig:flowchart}.

\vspace{2em}
\begin{table}[h]
\centering
    \begin{tabular}{|c|c|c|} 
 \multicolumn{1}{c}{} & \multicolumn{1}{c}{O3 online retractions} & \multicolumn{1}{c}{O3 online nonretracted alerts} \\ 
 \hline
Signal & $13/23$ & $52/56$ \\
\hline
Signal + Noise & $1/23$ & $3/56$ \\
\hline
Noise  & $9/23$ & $1/56$ \\
\hline
\end{tabular}

    \caption{Of the 23 O3 retracted public alerts, the QoQ test classified 9 as ``Noise" and 1 as ``Signal $+$ Noise". Out of the 56 O3 low latency astrophysical events, 3 were classified as ``Signal $+$ Noise", and 1 event was flagged as ``Noise". 
    A visual inspection supports this classification. \ref{appen_a} contains more details on these flagged retracted and O3 low latency events. }
    \label{tab_retrs_O3} 
\end{table}

\section{MDC Data and PyCBC background analysis}\label{mdc_pycbc_sec}

\subsection{MDC Data}
\begin{figure}[h]
    \centering
    \includegraphics[width=0.55\textwidth]{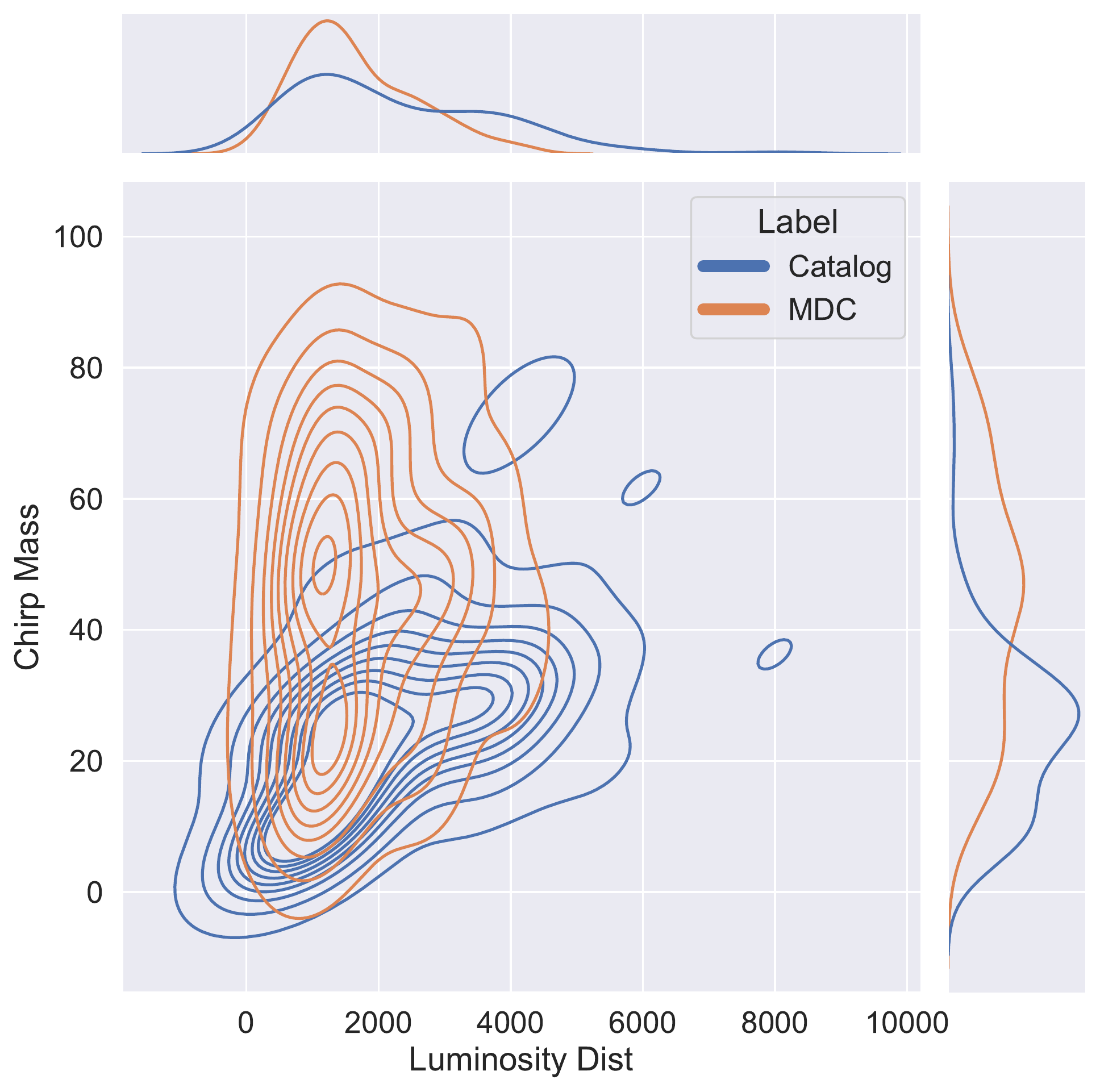}
    \caption{Joint plot distribution of the source chirp mass and source luminosity distance of the Catalog events and the MDC injections. The MDC set contains multiple injections at much lower luminosity distance and much higher chirp mass compared to expected values from an astrophysical population.}
    \label{fig:cmass_lumdist}
\end{figure}

To more robustly assess the QoQ test response to GW events, we analyze a set of BBH injections. We use the injection set from the O3 Mock Data Challenge (MDC). This MDC set contains signals injected in the O3 LIGO Livingston (LLO) and LIGO Hanford (LHO) data between Jan 1, 2020, and Feb 14, 2020. The dataset contains injections with source frame masses between 1 $M_\odot$ and 100 $M_\odot$, injected isotropically in the sky. The purpose of these injections is to evaluate the detection efficiency of the search pipelines, perform latency and alert generation checks as well as test source parameter estimation. The MDC injections will thus allow us to perform end-to-end testing of the low latency search pipelines in preparation for O4 \cite{MDC_5th, MDC_paper}.
Since the bigger goal of these MDC injections is to check pipeline consistency, the distribution of their source properties is not astrophysical. Compared to the catalog events, there is an excess of low luminosity distance, and high mass events, as shown in Fig \ref{fig:cmass_lumdist}. We will correct for this when we draw conclusions for our QoQ method, as ultimately, we need to assess the performance on the astrophysical population.

To isolate the injection and background data for binary black hole coalescences, we use the thresholds: $m_{1}>5$ $M_\odot$ and $m_{2}>5$ $M_\odot$ where $m_{1}$ and $m_{2}$ refers to primary and secondary mass with $m_{1}$ $>$ $m_{2}$. To ensure the high significance of events, we also use injection events with $\rho>5$ 
where $\rho$ represents the signal-to-noise ratio.
With these cuts, we get a total of 1354 MDC Injections.
Since the MDC waveforms were injected into the O3 data, some of them are polluted by the presence of very high SNR glitches near them. To enhance the purity of this dataset, we removed those injections that contained omicron transients with SNR above 50 within a 3-second window of the injection time. This reduced the size of the injection set marginally from 1354 to 1331 injections. Finally, we apply a threshold on network SNR of $\rho_{net} > 10$ to characterize found injections \cite{KAGRA:2021duu}. This brings the total number of mock injections to 1053. 
Next, as shown in Fig \ref{fig:cmass_lumdist} the MDC injection set consists of high mass and low luminosity distance events and does not match the expected astrophysical distribution. To obtain a set that more closely resembles an astrophysical population, we use Importance Sampling described next.

\subsection{Importance Sampling}
Say we would like to calculate the expectation value of a random variable f(x), where x is a random vector. Assuming continuous distribution and p(x) the probability density, we can write the expectation value as: 
\begin{equation}
    \mu_{f} = \int f(x)p(x) dx
\end{equation}

Now, let’s say we have another distribution defined by the probability density $q(x)$, and we would like to calculate $\mu_{f}$ by drawing samples from this new distribution $q(x)$. Given that $q(x) \neq 0$ whenever $f(x)p(x)\neq 0$, we can take the above equation and modify it as:
\begin{equation}
    \mu_{f} = \int \frac{f(x) p(x)}{q(x)} q(x) dx
\end{equation}

The ratio $\frac{p(x)}{q(x)}$ weights the samples from this new distribution. Certain elements from q(x) have a higher impact or they are more ``important” and the weights emphasize these elements while sampling.  Essentially, this allows to calculate the expectation value of samples drawn from q(x) by adjusting their importance with respect to the original distribution p(x) \cite{Tom:2016}. So for samples $X_{1}....X{n}$ the importance sampling algorithm can be written as:
\begin{equation}
    \hat{\mu} = \frac{1}{n}\sum_{i=1}^{n} \frac{f(X_{i})p(X_{i})}{q(X_{i})}
\end{equation}

In our case, the $p(x)$ and $q(x)$ refers to the Catalog and MDC Injections respectively and $f(x)$ is the QoQ test function as shown in Fig \ref{fig:flowchart}. We use luminosity distance and source chirp mass to generate the kernel density estimate, and so we can rewrite the above equation as:

\begin{equation}
    \hat{\mu} = \frac{1}{n}\sum_{i=1}^{n} \frac{f_{QoQ}(X_{i}, Y_{i})p_{cat}(X_{i}, Y_{i})}{q_{mdc}(X_{i}, Y_{i})}
\end{equation}

An MDC injection sample whose luminosity distance and source chirp mass closely resembles the respective distributions of Catalog events is assigned a higher weight as opposed to injections that do not. This way, we can sample all the MDC injections and calculate the weight-adjusted expectation value.

To calculate the density estimate, we use a Gaussian kernel:

\begin{equation}
    K(x,y) = \frac{1}{2\pi w_{x}w_{y}} \frac{1}{n}\sum_{i=1}^{n} exp\left[-\frac{(x - x_{i})^2}{2w_{x}^2} - \frac{(y - y_{i})^2}{2w_{y}^2}\right]
\end{equation}
where $w_{x}$ and $w_{y}$ are the bandwidth values. 

\subsection{Q-occupancy analysis of BBH data}
Each event follows the steps outlined in the flowchart shown in Fig \ref{fig:flowchart}. For each  MDC injection, we first calculate its weight $\frac{p_{cat}(X_{i}, Y_{i})}{q_{mdc}(X_{i}, Y_{i})}$. We then apply the QoQ test on this event as outlined in Fig \ref{fig:flowchart}.
Any event that is not flagged by the nonoverlap method is classified as a ``Signal". 
Any event that fails to pass the nonoverlap method then goes through the overlap method. If the event is flagged at this stage, it is then classified as ``Noise", otherwise, it is classified as ``Signal $+$ Noise". We use a total of three sets of QoQ thresholds, Thresh A, Thresh B and Thresh C. The non overlap and overlap Q occupancy threshold values for Thresh A, Thresh B and Thresh C are ($[0,0,2,0,0]$,$[2,1,1]$), ($[0,0,3,0,0]$,$[3,1,1]$) and ($[0,0,4,0,0]$,$[4,1,1]$) respectively. The fraction of events flagged at each stage of  Fig \ref{fig:flowchart}, for each set of QoQ thresholds is shown in Table \ref{tab:tab_nonovlp_bbh}.

Ideally, we would not misclassify a true gravitational-wave signal as noise. There are multiple reasons why a small percentage of MDC injections are classified as noise. These chirp signals were injected into the actual O3 data from LLO and LHO. And so, a number of these signals are polluted by the presence of noise transients near them. The Q-transforms of the injection events that are flagged by the non-overlap method but not by the overlap method reveal the presence of nearby transient noise along with the chirp morphology.

\vspace{1cm}
\begin{table}[h]
\centering

\setlength{\tabcolsep}{5pt}
\begin{tabular}{l r r r r r r}
\hline
\hline
\multicolumn{1}{c}{} & \multicolumn{3}{c}{Injections} &  \multicolumn{3}{c}{Background} \\
\cline{2-4} \cline{5-7}
 \multicolumn{1}{c}{} & \multicolumn{1}{l}{Thresh A} & \multicolumn{1}{l}{Thresh B} & \multicolumn{1}{l}{Thresh C} & \multicolumn{1}{r}{Thresh A} & \multicolumn{1}{r}{Thresh B}  & \multicolumn{1}{r}{Thresh C} \\ 
 \hline\hline
Signal & $95^{100.0}_{88.9}$\% & $95.6^{100.0}_{89.5}$\% & $96.2^{100.0}_{90.2}$\%  & $52.7^{55.1}_{50.2}$\% & $55.1^{57.5}_{52.6}$\% & $57.2^{59.6}_{54.7}$\% \\
\hline
Signal + Noise & $2.9^{3.7}_{2.0}$\% & $2.3^{3.2}_{1.4}$\% & $1.7^{2.6}_{0.8}$\% & $4.2^{5.2}_{3.2}$\% & $4.1^{5.1}_{3.1}$\% & $4.1^{5.1}_{3.1}$\%\\
\hline
Noise & $2.1^{2.7}_{1.5}$\% & $2.1^{2.7}_{1.5}$\% & $2.1^{2.7}_{1.5}$\%  & $43.1^{45.5}_{40.6}$\% & $40.8^{43.2}_{38.4}$\% & $38.7^{41.0}_{36.3}$\%\\
\hline
\hline
 \end{tabular}
    \caption{ This table shows the results on MDC Injections and PyCBC background data for pixel-energy threshold of 60 and QoQ threshold values of 2, 3 and 4 (Thresh A, Thresh B and Thresh C respectively). For QoQ threhsold of 2, the test classifies 95$\%$ of the Injection data as ``Signal", 2.9 $\%$ as ``Signal $+$ Noise" and 2.1 $\%$ as just ``Signal". The lower and upper limits cover the 95 $\%$ confidence interval for these values. Figure \ref{fig:inj_back_roc} shows these results for other values of pixel-energy thresholds.
    }
    \label{tab:tab_nonovlp_bbh} 
\end{table}

\begin{figure}[h]
    \centering
    \includegraphics[width=0.65\textwidth]{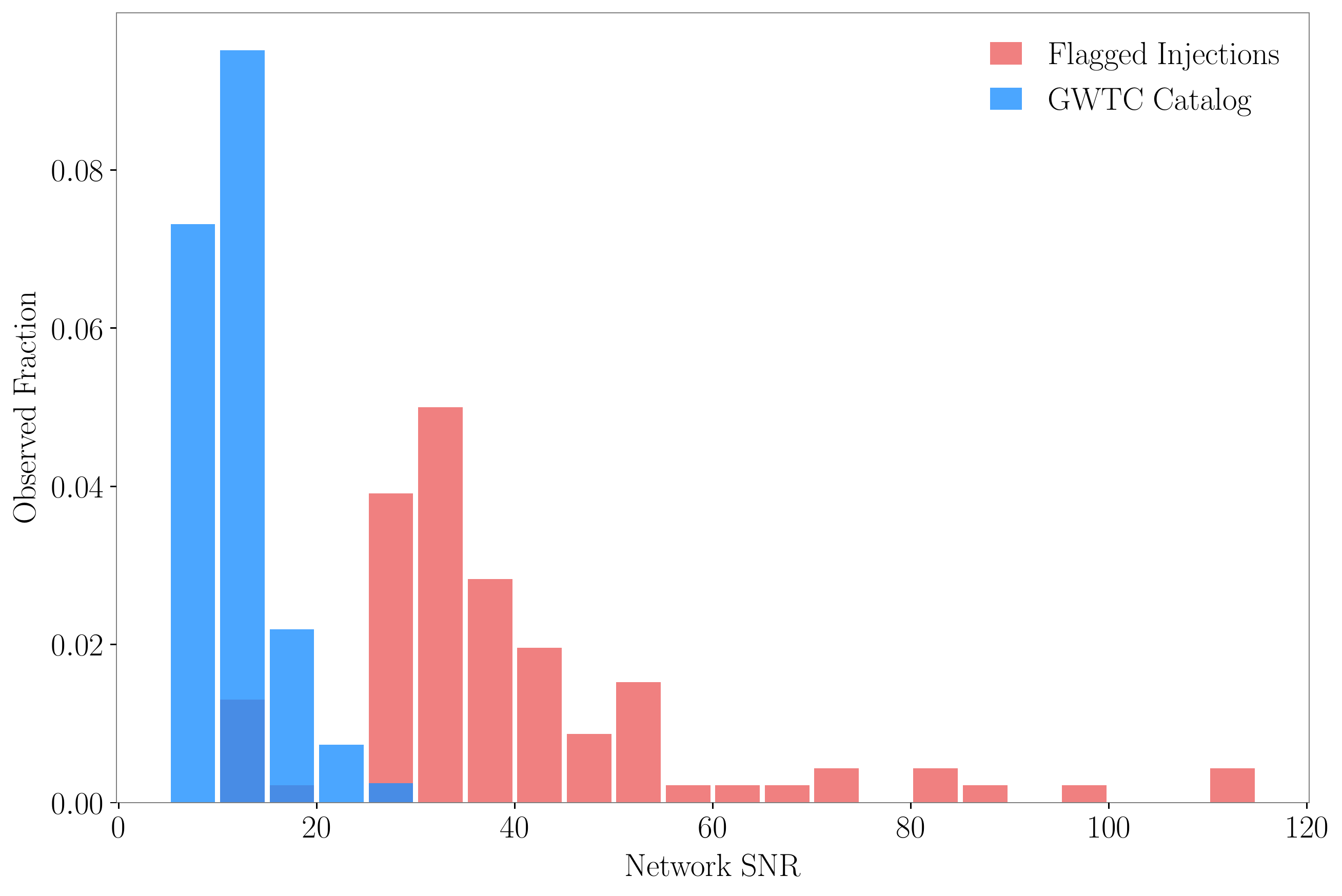}
    \caption{About $2 \%$ of the MDC injections were classified as ``Noise" by the QoQ test. This figure shows the network SNR distribution comparison between these injections flagged as ``Noise" and Catalog events. The histogram is normalized so that the area under each distribution sums to 1. These flagged injections have a much high network SNR, not expected from astrophysical signals.}
    \label{fig:snr_dist_hist}
\end{figure}

\begin{figure}
  \begin{subfigure}[t]{.48\textwidth}
    \centering
    \includegraphics[width=\linewidth]{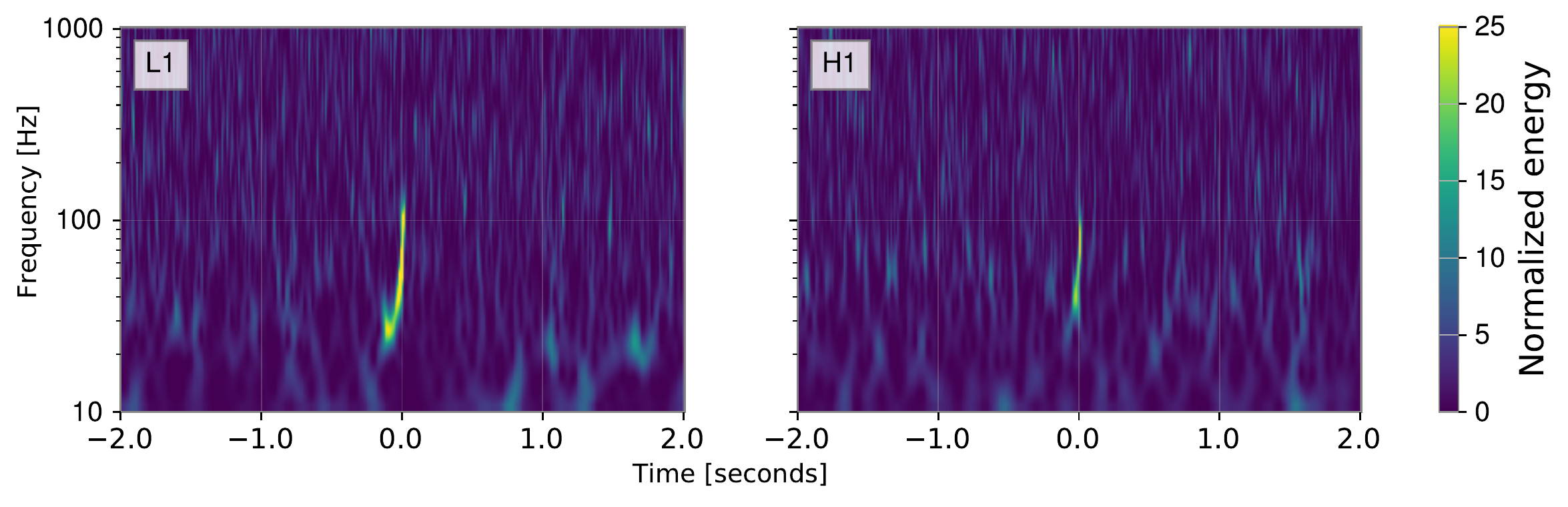}
  \end{subfigure}
\hspace{1cm}
  \begin{subfigure}[t]{.48\textwidth}
    \centering
    \includegraphics[width=\linewidth]{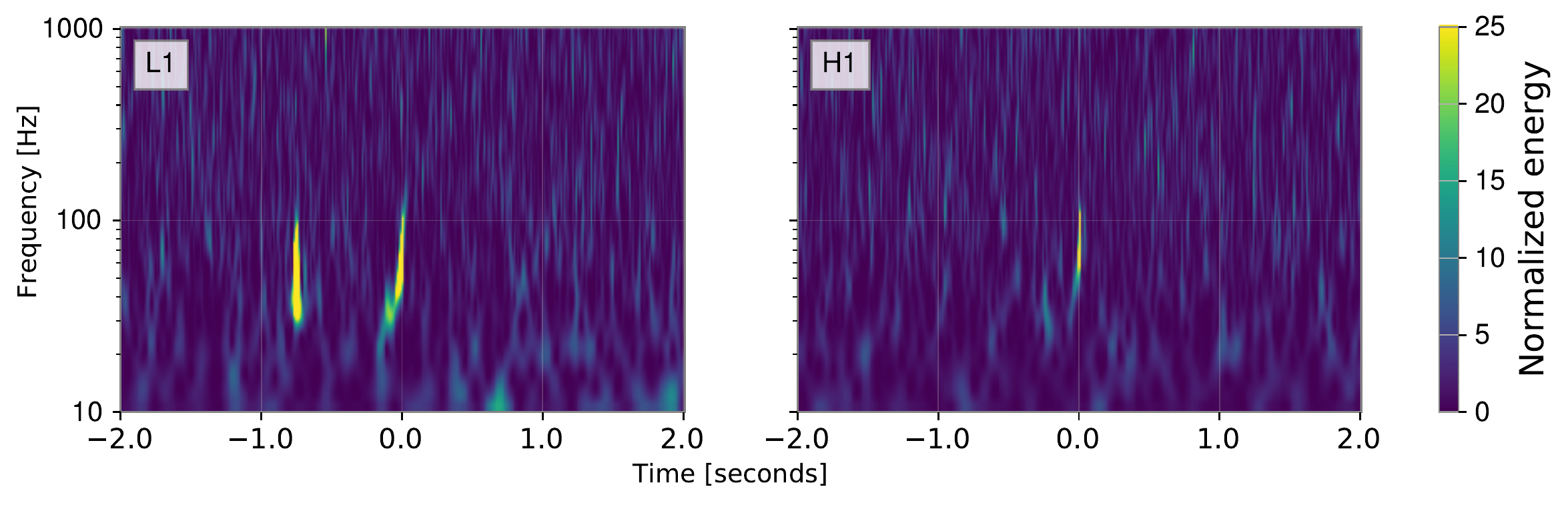}
  \end{subfigure}

  \vspace{0.5em}
  \begin{subfigure}[t]{.48\textwidth}
    \centering
    \includegraphics[width=\linewidth]{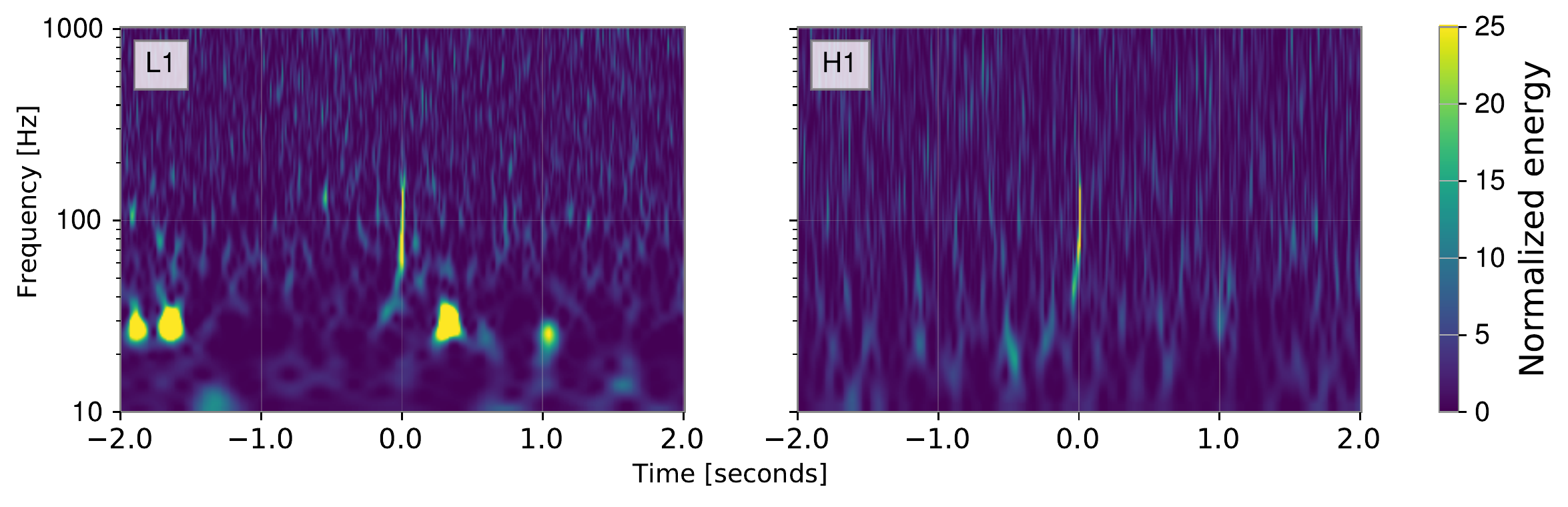}
  \end{subfigure}
  \hspace{1.1cm}
  \begin{subfigure}[t]{.48\textwidth}
    \centering
    \includegraphics[width=\linewidth]{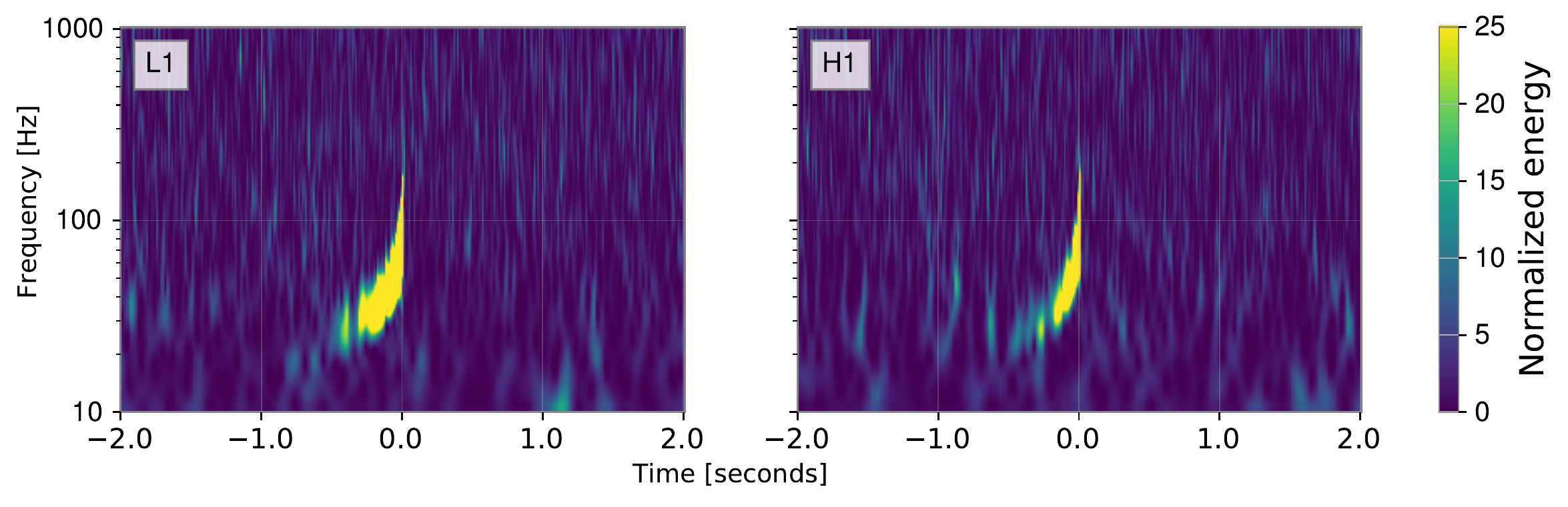}
  \end{subfigure}
\caption{Q-transforms of BBH injections analyzed at different stages in the Fig \ref{fig:flowchart}.\emph{Top left}: An MDC injection classified as ``Signal" by the QoQ test. This event is not flagged by the non overlapping analysis. About $95 \%$ of the MDC injections are classified as ``Signal" by the QoQ test. \emph{Top right} and \emph{Bottom left}: These two injections are flagged by non overlapping analysis but not by overlapping, and are classified as ``Signal  $+$ Noise". For both of these injections, we can observe the presence of transients in the L1 data. 
About $3 \%$ of the MDC injections are classified in this category by the QoQ test. \emph{Bottom right}: This injection is classified as noise due to its very high energy content. About $2 \%$ of the MDC injections were classified as ``Noise" by the QoQ test.  }\label{qscans_eachstep}
\end{figure}

For the Thresh A case, about $2.0\%$ of the injection events are classified as noise. Since these events come from the injection set, this fraction represents the false positive in our analysis. Fig \ref{fig:snr_dist_hist} compares the Network SNR of these flagged injections with the GWTC Catalog. As we can see from this comparison, these flagged MDC injections are very loud, and their SNR distribution does not resemble the distribution expected from a set of astrophysical events \cite{LIGOScientific:2020kqk}. The importance sampling weight of all these events is below 1.
Furthermore, Fig \ref{qscans_eachstep} shows the Q-transforms of the three categories of injections flagged at different stages of the QoQ test (``Signal", ``Signal $+$ Noise", and ``Noise").

\subsection{PyCBC Background}
In order to assess the QoQ's test ability to reject false pipeline alerts, we analyze a set of background events generated by PyCBC \cite{Davies:2020tsx,Usman:2015kfa}. We use the PyCBC-broad analysis from GWTC PyCBC is a matched filtering pipeline that identifies triggers by finding peaks in the matched filter SNR between the data and a template bank of waveforms. Events are then constructed by matching coincident sets of these triggers in the detector network. Each event is compared to a background to assess the false alarm rate. 

Background events are constructed through time shifts, where the triggers from one detector are shifted in time by more than the gravitational-wave travel time compared to other detectors, breaking the astrophysical coincidence. The background events we use come from the PyCBC-broad analysis of the first half of the third LVK observing run, O3a. These analyses were used to produce results in GWTC-2.1~\cite{LIGOScientific:2021usb}. We use the PyCBC `exclusive' background only, meaning any triggers which appear in the un-shifted coincident sets of triggers are not used, ensuring the background is not contaminated by potential astrophysical signals. These background events mimic possible false coincidences that could appear during an online observing run, providing a useful data set to evaluate the QoQ test.

To obtain the background data for binary black hole coalescences, we use the thresholds: $m_{1}>5$ $M_\odot$ and $m_{2}>5$ $M_\odot$ where $m_{1}$ and $m_{2}$ refers to primary and secondary mass with $m_{1}$ $>$ $m_{2}$. And similar to MDC injections, to ensure the high significance of events, we use background events with $ifar>50$ years where ifar is the inverse false alarm rate. 
With these cuts, we get a total of 1591 PyCBC background coincidences. The importance sampling is not required for the background data since they are not astrophysical events and are not expected to match the astrophysical distribution of source properties.

\subsubsection{Analysis of BBH Background flagged as noise}
\begin{figure}[h]
    \centering
    \includegraphics[width=0.65\textwidth]{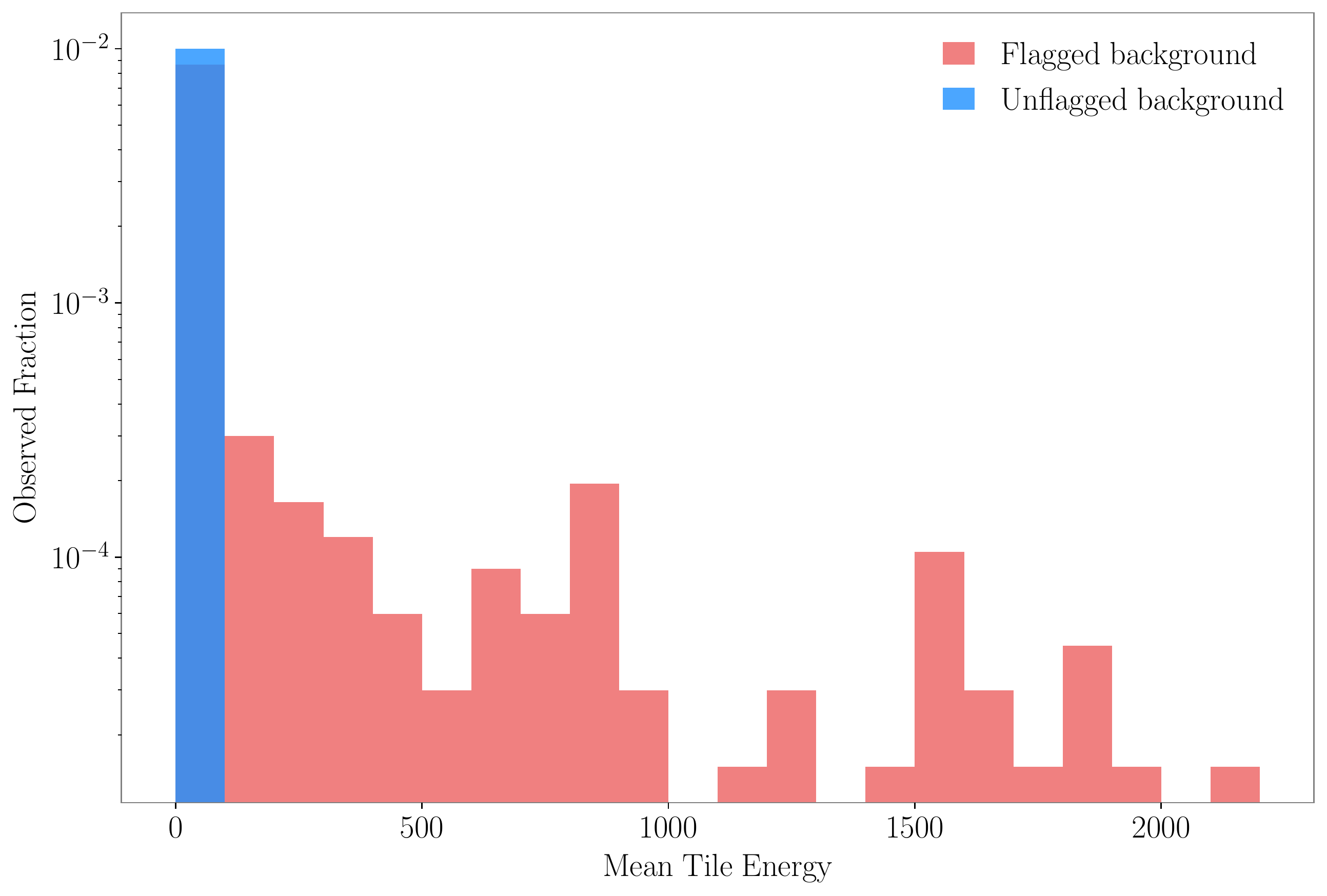}
    \caption{A comparison of distribution of mean pixel energies between unflagged background events and background events flagged as noise by the QoQ analysis. }
    \label{fig:bkg_mean_tile}
\end{figure}

Depending on the Q-occupancy threshold, Table \ref{tab:tab_nonovlp_bbh} shows that the test flags between $39\%$ and $43\%$ of the background events as noise, while between $53 \%$ and $57\%$ are classified as ``Signals". As mentioned earlier in Section \ref{retr_events_sec}, O3 retractions events fall into two main categories: events with a clear presence of transient noise and low SNR events with no visible transient noise present. The majority of the background coincidences not flagged by the Q-occupancy test belong to this latter category of events. Fig \ref{fig:bkg_mean_tile} shows pixel-energy comparison of flagged and unflagged background coincidence. 
The Q scans of these unflagged background coincidences do not contain considerable noise above the pixel-energy threshold of 60 and thus were not picked up by the Q-occupancy test.
However, since the background events have $ifar>50$ years, these are considered highly significant events by the PyCBC search pipeline. With a Q-occupancy threshold of 2 and pixel-energy threshold of 60, the QoQ test flagged $43 \%$ of BBH background as ``Noise". This is very consistent with the amount of O3 retractions classified as ``Noise" by the QoQ test, as shown in Table \ref{tab_retrs_O3}. This suggests that we can expect QoQ test to flag similar percentage of false positives in O4.

\begin{figure}[h]
\captionsetup[subfigure]{}
   \centering
    \begin{subfigure}{0.48\textwidth}
        \centering
         \includegraphics[width=0.9\textwidth, height=0.60\textwidth]{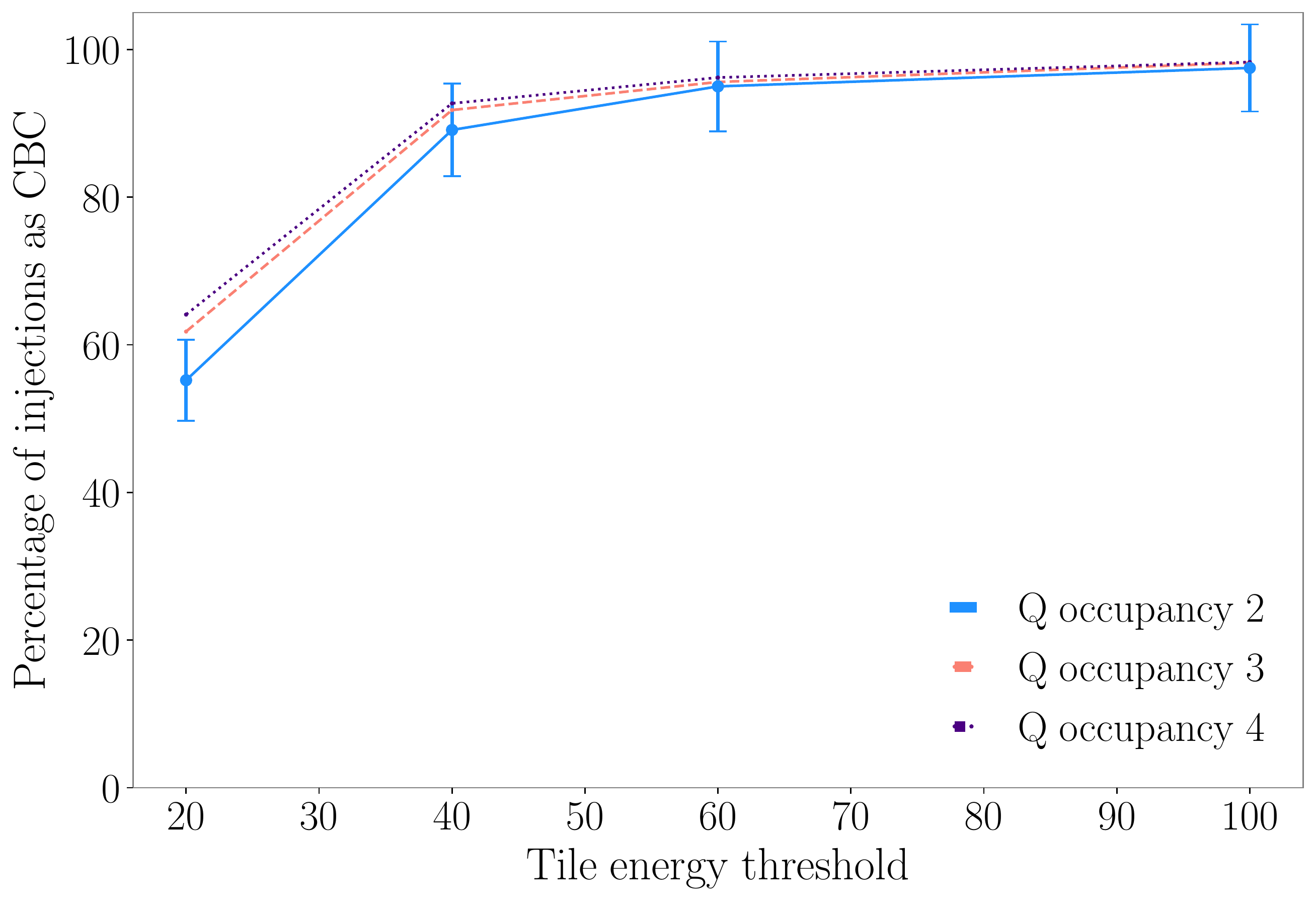}
         \label{fig:inj_roc}
    \end{subfigure}
    \begin{subfigure}{0.48\textwidth}
        \centering
         \includegraphics[width =0.9\textwidth, height=0.60\textwidth]{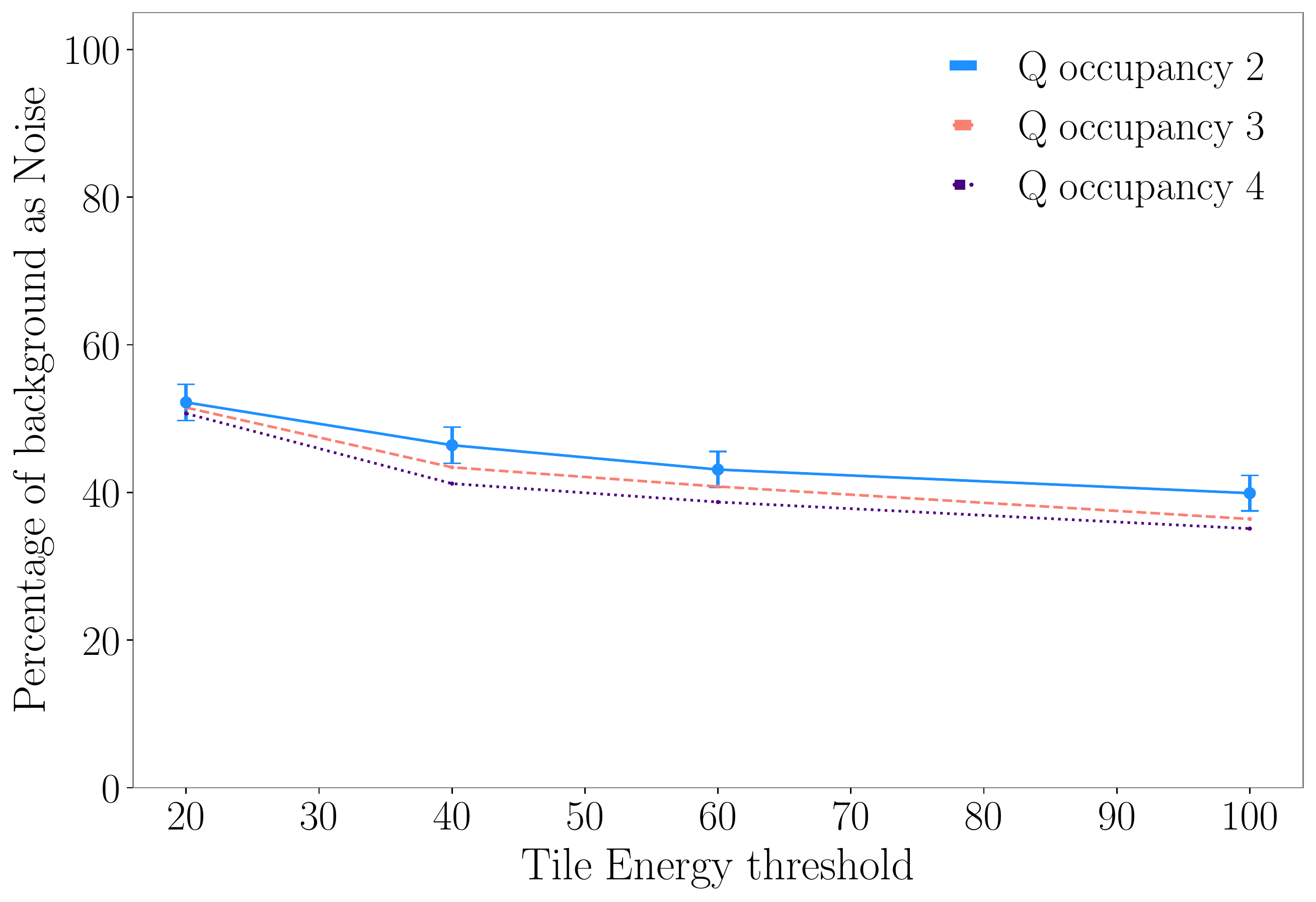}
         \label{fig:bkg_roc}
    \end{subfigure}
    \caption{Here we show the variation in MDC Injections classified as ``CBC" and PyCBC background classified as ``Noise" as we change the pixel-energy threshold and Q-occupancy threshold values. The 95 $\%$ confidence interval error bars are shown for Q-occupancy threshold of 2. \emph{Left}: As we increase the pixel-energy threshold, a larger fraction of injections are classified as ``CBC". However, after pixel-energy threshold of 60, this change is rather small. \emph{Right}: As expected, increasing pixel-energy threshold results in smaller percentage of background events classified as ``Noise".}
    \label{fig:inj_back_roc}
\end{figure}



\begin{table}[h]
\centering
    \begin{tabular}{|c|c|c|c|} 
 \multicolumn{1}{c}{} & \multicolumn{1}{c}{O3 online retractions} &
 \multicolumn{1}{c}{O3 online non retracted} &
 \multicolumn{1}{c}{O3 Catalog events} \\ 
 \hline
Signal & $13/23$ & $52/56$ & $67/74$ \\
\hline
Signal + Noise & $1/23$ & $3/56$ & $7/74$ \\
\hline
Noise  & $9/23$ & $1/56$ & $0/74$ \\
\hline
\end{tabular}
    \caption{ In this Table, we show the results of QoQ classification on O3 events that include offline and online events and on O3 Catalog events, along with the retracted events.
    \ref{appen_a} contains more details on the O3 events classified as Signal $+$ Noise, and as Noise by the QoQ test. }
    \label{tab_retrs_O3all} 
\end{table}

\section{O3 offline analysis}\label{O3_sec}
In Table \ref{tab_retrs_O3} 
we compare the results of the QoQ test on O3 retracted events with non-retracted events found in low latency. Now we extend this comparison to include the events in O3 Catalog. As opposed to online analyses, offline analyses use cleaner and better-calibrated data over which the search is performed. This leads to offline events not found in online analyses since more computationally extensive methods can be used on an improved dataset. Also, some of the events found in low latency may not pass the thresholds of FAR and/or p-astro and thus do not appear in the Catalog list.
The catalog events list contains a total of 74 candidates, 44 of which are found in low latency and 30 in offline analyses \cite{LIGOScientific:2021usb,LIGOScientific:2021djp}. Tab \ref{tab_retrs_O3all} shows the results on this full set of events from O3. 
Out of the 74 events in O3 catalog list, the QoQ test classifies 7 of these events as Signal with presence of transient noise, while none of the events are classified as just noise. 
In the Q scans plotted in \ref{fig:figure_appendixO3} we can see the presence of transient noise near the candidate time for each of these 7 events.
Astrophysical candidates polluted with nearby transient noise require a more careful event validation to rule out their instrumental origin \cite{Davis:2021ecd}.


\section{Discussion}\label{disc_sec}
The presence of short-duration transient noise adversely impacts the quality of gravitational-wave strain data. 
The origin of these transients is in the complicated web of detector hardware, and it is often challenging to find their exact source. Apart from masking the GW signal morphology, the non-Gaussian transient noise may sometimes mimic the astrophysical signals and end up as search pipeline detection candidates.
It is thus essential to build the ability to identify and characterize potential noise artifacts in real time so that false public alerts can be reduced.
Our QoQ method approaches the classification problem from first principles on how excess power in a time-frequency decomposition is expected to appear for astrophysical signals vs noise and derives straightforward criteria to separate events into three (mutually exclusive) categories: signal, noise, and signal plus noise. 
Few analysis parameters allow the method to tune to specific astrophysical sources and/or noise artifacts.
The application of the method we described here is within the context of binary black hole candidate events as they are identified by matched filtering (or otherwise) detection algorithms.
Our analysis starts with the Q-transformation of the strain data and thresholds on the signal energy of the time-frequency pixels. Then it uses the pixel occupancy over several time-frequency windows in order to perform the classification problem. We tested QoQ extensively with both simulated GW waveforms the LIGO-Virgo-KAGRA Collaborations introduced as part of testing the low-latency alert infrastructure~\cite{MDC_5th} as well as background data from the search for binary black holes with PyCBC in O3~\cite{Davies:2020tsx}.
This allowed us to explore various options for pixel energy thresholds and actual pixel occupancy ones, leading to ROC curves.

The application of our method in the above dataset was mostly geared in minimizing the false dismissal of GW-like events while maximizing the rejection of noise.
While there is room for further optimization, our present choice of analysis thresholds leads to over 40\% reduction of the background (noise) while misidentifying as noise about 2\% of GW-like signals.
These numbers are consistent with what we obtain when applying QoQ on actual public alerts that were issued by the LIGO-Virgo-KAGRA collaboration in real-time during O3 as well as with events in the O3 catalog as published by the LIGO-Virgo-KAGRA collaborations \cite{LIGOScientific:2021djp}.
The ability to reject additional noise events with QoQ (and beyond the efficacy we currently have) faces challenges as about half of the noise and/or background events show little signal energy (Signal-to-Noise-Ratios below 10) in their time-frequency decomposition.

It is an efficient algorithm to run in real-time and integrate into workflows for astronomical alert generation, mostly in the form of preventing false alerts from dissemination.
Additionally, its ability to identify the simultaneous presence of signal and noise within an analysis window can facilitate the workflow of further denoising/deglitching as well as follow-up GW parameter estimation analyses.
The occurrence of transient noise (glitches) may introduce biases in the parameters estimated for a given GW candidate \cite{Pankow:2018qpo, Payne:2022spz, Macas:2022afm, Kwok:2021zny, LIGOScientific:2021djp}. The ability to quantify and identify automatically such types of events will further assist their analysis with parameter estimation techniques.
The enhanced sensitivity of the GW detectors in O4 will result in an increased rate of both the GW candidates and likely new sources of noise transients. Tools such as QoQ will be helpful in improving the purity of resulting real-time alerts.



\ack{}
The authors would like to thank the low latency and detector characterization working groups of the LIGO Scientific Collaboration for feedback while carrying out this work. We would especially like to thank Geoffrey Mo and Deep Chatterjee for their input. SS, EM and EK acknowledge support from the United States National Science Foundation (NSF) under award PHY-1764464 to the LIGO Laboratory and OAC-2117997 to the A3D3 Institute. GCD acknowledges the Science and Technology Funding Council (STFC) for funding through grant ST/T000333/1.
M.~W.~Coughlin acknowledges NSF support under awards PHY-2010970 and OAC-2117997. S. Ghosh acknowledges NSF support under award PHY-2110576.
R. Essick is supported by the Natural Sciences \& Engineering Research Council of Canada (NSERC).

This material is based upon work supported by NSF's LIGO Laboratory which is a major facility fully funded by the National Science Foundation.
This research has made use of data and software obtained from GWOSC (\href{http://www.gw-openscience.org/}{gw-openscience.org}), a service of LIGO Laboratory, the LIGO Scientific Collaboration, the Virgo Collaboration, and KAGRA Collaboration.
The authors gratefully acknowledge the support of the US NSF for the construction and operation of the LIGO Laboratory and Advanced LIGO as well as STFC of the United Kingdom, and the Max-Planck-Society for support of the construction of Advanced LIGO. 
Additional support for Advanced LIGO was provided by the Australian Research Council. 
Advanced LIGO was built under award PHY-0823459.
LIGO was constructed by the California Institute of Technology and Massachusetts Institute of Technology with funding from the National Science Foundation and operates under Cooperative Agreement PHY-1764464. The authors are grateful for computational resources provided by the LIGO Laboratory and supported by National Science Foundation Grants PHY-0757058 and PHY-0823459.

\newpage
\appendix
\section{QoQ analysis of O3 Catalog  and retracted events}\label{appen_a}

\vspace{2em}
\begin{table}[h]
\centering
\begin{tabular}{|c|c|c|c|c|c|c|c|c|} 
\hline
 \multicolumn{1}{c}{IFO} & \multicolumn{1}{c}{Freq band [Hz]} & \multicolumn{1}{c}{t1} & \multicolumn{1}{c}{t2} & \multicolumn{1}{c}{t3} & \multicolumn{1}{c}{t4} & \multicolumn{1}{c}{t5} & \multicolumn{1}{c}{QoQ Flag} & \multicolumn{1}{c}{Superevent}\\ 
 \hline
L1 & $10-100 $ & $42.0$ &  $78.7$ & $19.8$	&  $10.8$ &  $0.0$ & Noise & S200308e \cite{GCN27347}\\
\hline
L1 & $10-100 $ & $0.0$ &  $0.0$ & $17.2$ &  $5.1$ &  $0.0$ & Noise & S200108v \cite{GCN26665}\\
\hline
L1 & $10-100 $ & $0.0$ &  $0.0$ & $0.0$	&  $1.4$ &  $5.1$  & Noise & S190405ar \cite{GCN24109}\\
\hline
H1 & $10-100 $ & $0.0$ &  $0.6$ & $2.3$	&  $0.0$ &  $0.0$  & Noise & S191212q \cite{GCN26394}\\
\hline
H1 & $10-100 $ & 24.0 & 	34.8&	24.6&	0.0	& 0.0  & Noise & S200106av \cite{GCN26641}\\
\hline
L1 & $10-100 $ & 7.1 &	6.0& 1.7&  0.0 & 0.0  & Noise & S200106au \cite{GCN26641}\\
\hline
L1 & $10-100 $ & 0.0 &	23.3&	22.0&	0.0&	0.0  & Noise & S191117j \cite{GCN26254}\\
\hline
L1 & $10-100 $ & 0.0 &	24.6 &	0.2 &	0.0 &	0.0   & Noise & S191120at \cite{GCN26265}\\
\hline
L1 & $10-100 $ & 0.0 &	0.0 &	2.5 &	0.0 &	0.0	  & Noise & S200116ah \cite{GCN26784}\\
\hline
L1 & $10-100 $ & 0.0 &	0.3 &	1.1 &	0.0 &	0.0	  & Signal $+$ Noise & S191220af \cite{GCN26512}\\
\hline

\end{tabular}

    \caption{QoQ Non Overlapping method analysis of the retracted events. The table shows the pixel occupancy values of the flagged retracted events. The first nine events in this table were also flagged by the Overlapping method, as shown in Fig \ref{fig:flowchart} and thus  are classified as ``Noise" by the QoQ test. The last event S191220af was only flagged by the Non-Overlapping method and is thus classifed as ``Signal + Noise".}
    \label{tab:O3retrpixvalsovlp} 
\end{table}

\begin{figure}
\centering
\begin{subfigure}{0.45\textwidth}
    \includegraphics[width=\textwidth]{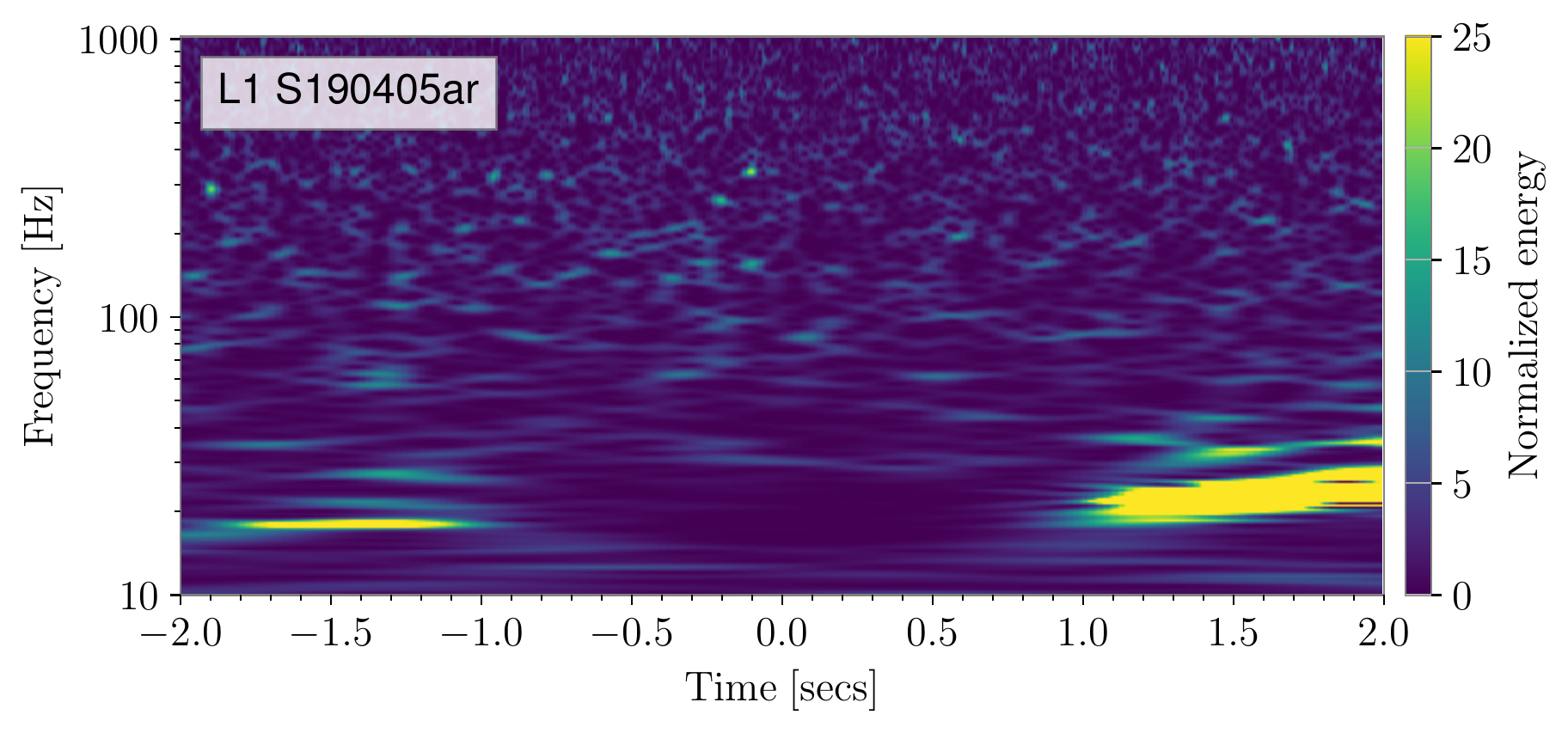}
    \caption{S190405ar}
\end{subfigure}
\hfill
\begin{subfigure}{0.45\textwidth}
    \includegraphics[width=\textwidth]{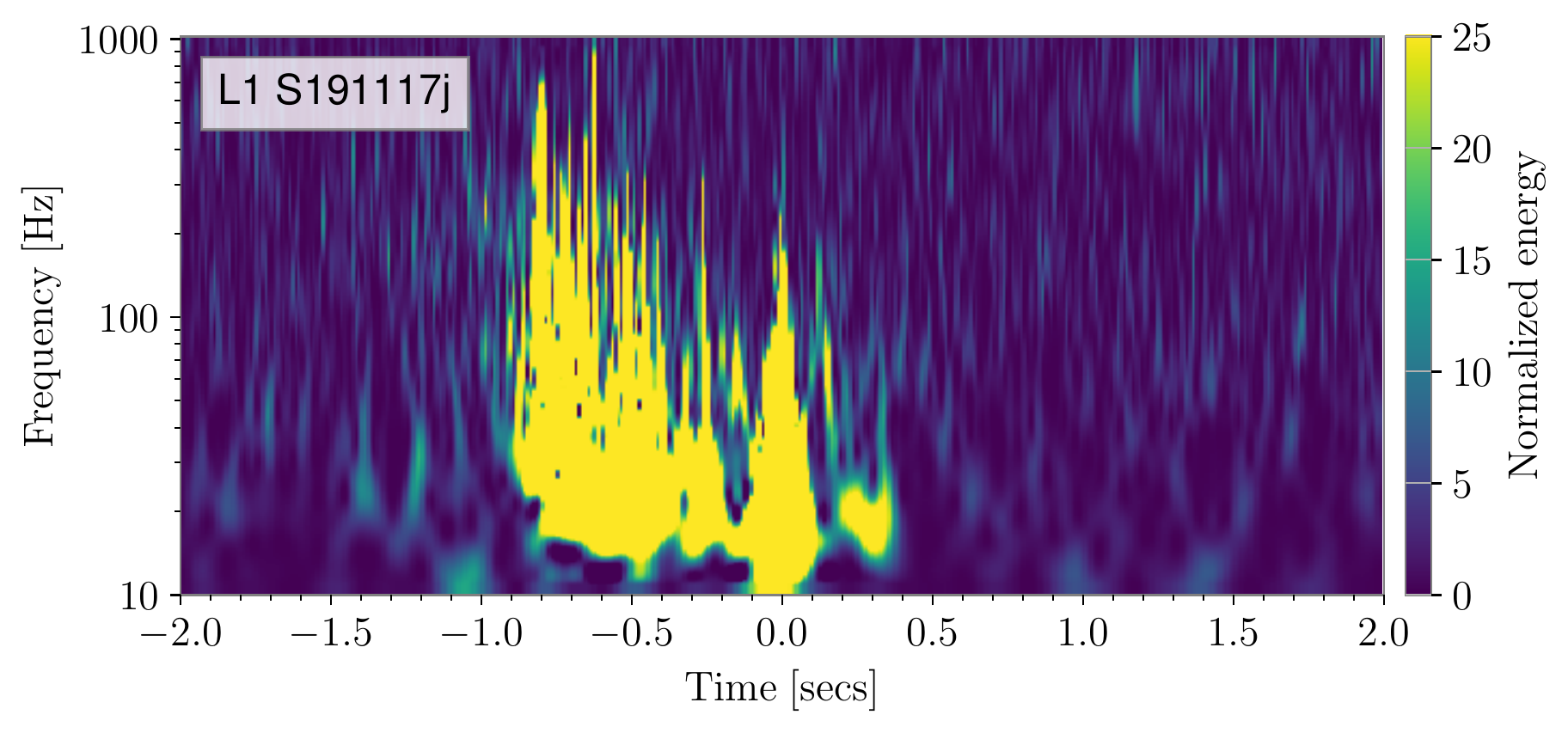}
    \caption{S191117j}
\end{subfigure}

\begin{subfigure}{0.45\textwidth}
    \includegraphics[width=\textwidth]{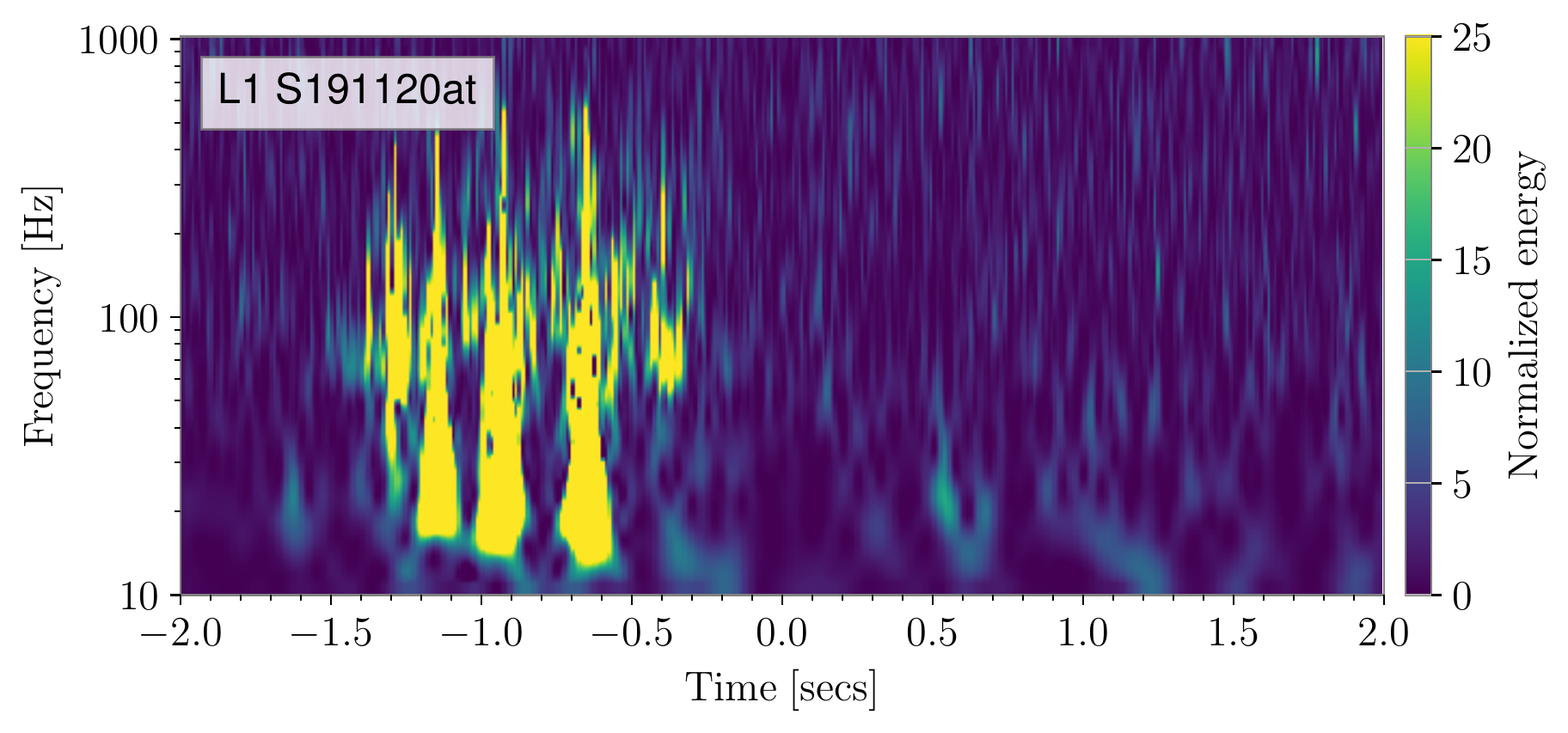}
    \caption{S191120at}
\end{subfigure}
\hfill
\begin{subfigure}{0.45\textwidth}
    \includegraphics[width=\textwidth]{O3retrflagged_1260174466_H1.pdf}
    \caption{S191212q}
\end{subfigure}

\begin{subfigure}{0.45\textwidth}
    \includegraphics[width=\textwidth]{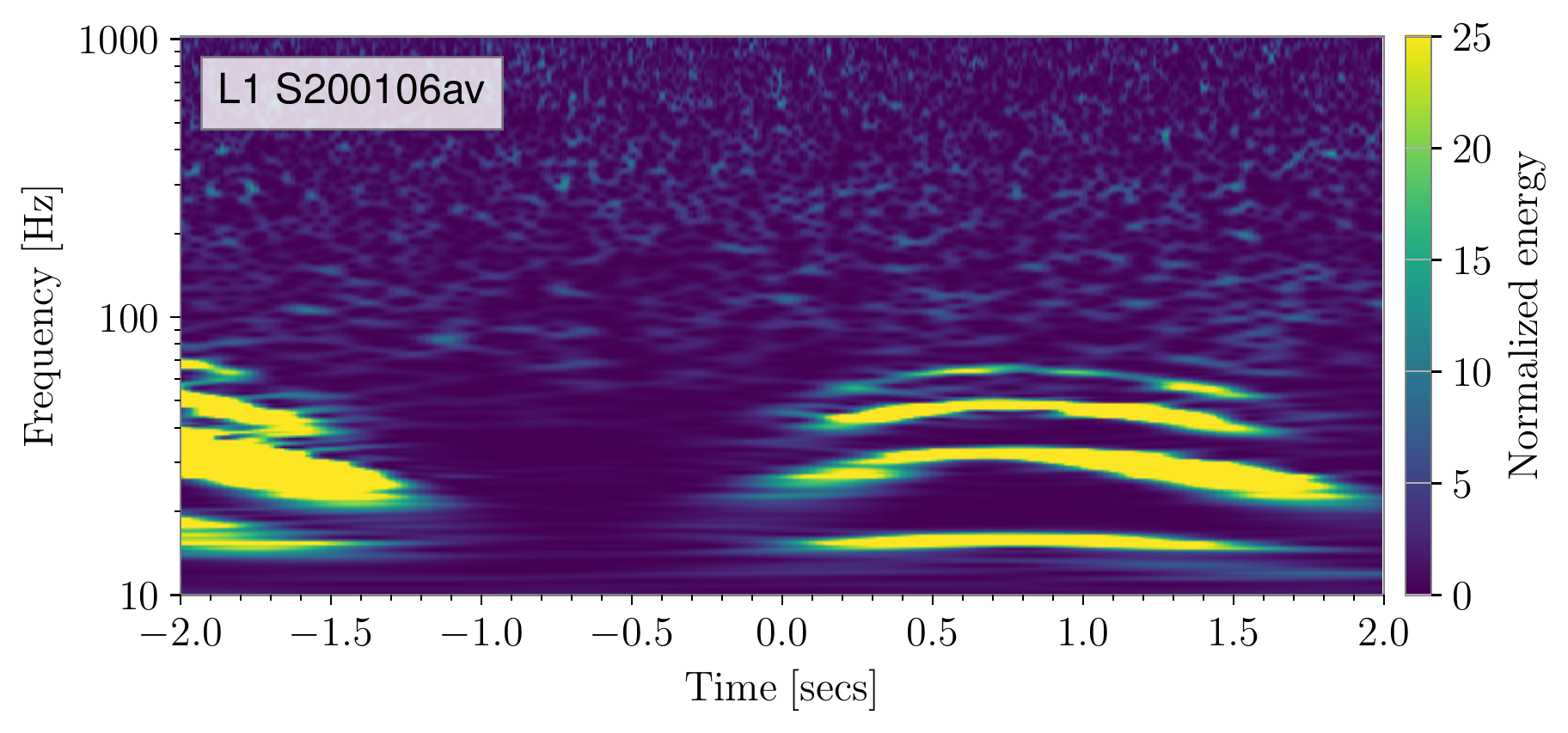}
    \caption{S19200106av}
\end{subfigure}
\hfill
\begin{subfigure}{0.45\textwidth}
    \includegraphics[width=\textwidth]{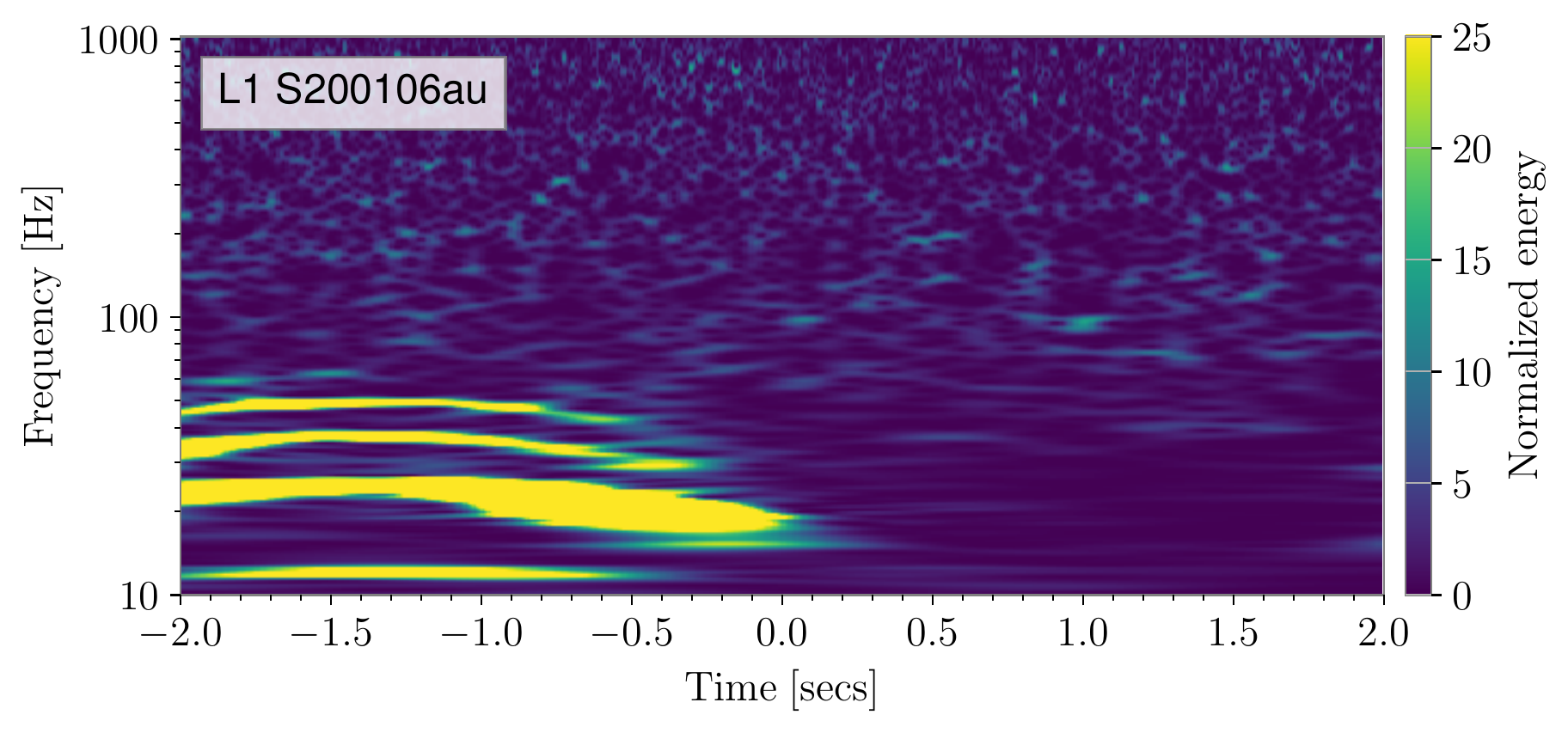}
    \caption{S200106au}
\end{subfigure}

\begin{subfigure}{0.45\textwidth}
    \includegraphics[width=\textwidth]{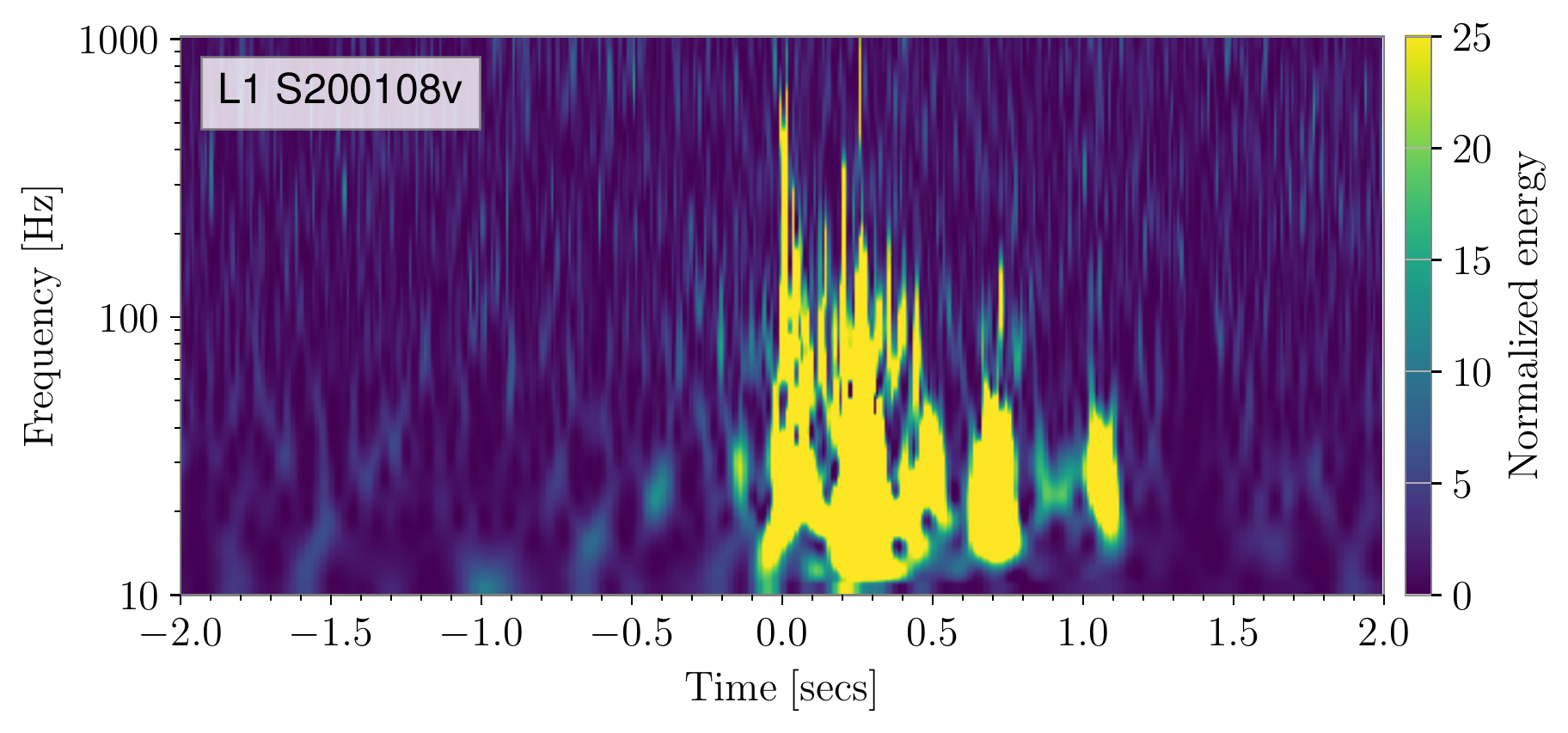}
    \caption{S200108v}
    \label{fig:first}
\end{subfigure}
\hfill
\begin{subfigure}{0.45\textwidth}
    \includegraphics[width=\textwidth]{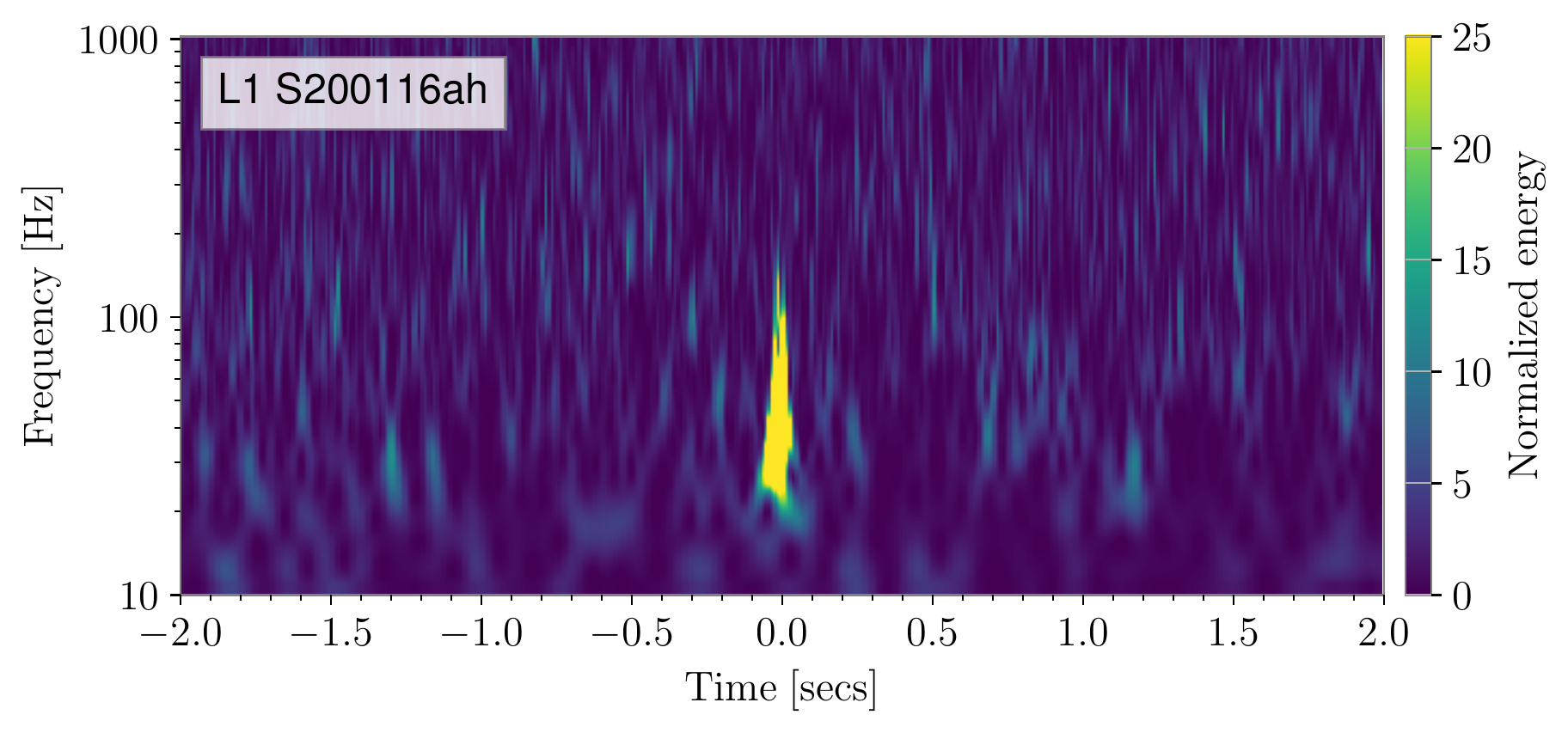}
    \caption{S200116ah}
    \label{fig:second}
\end{subfigure}


\begin{subfigure}{0.45\textwidth}
    \includegraphics[width=\textwidth]{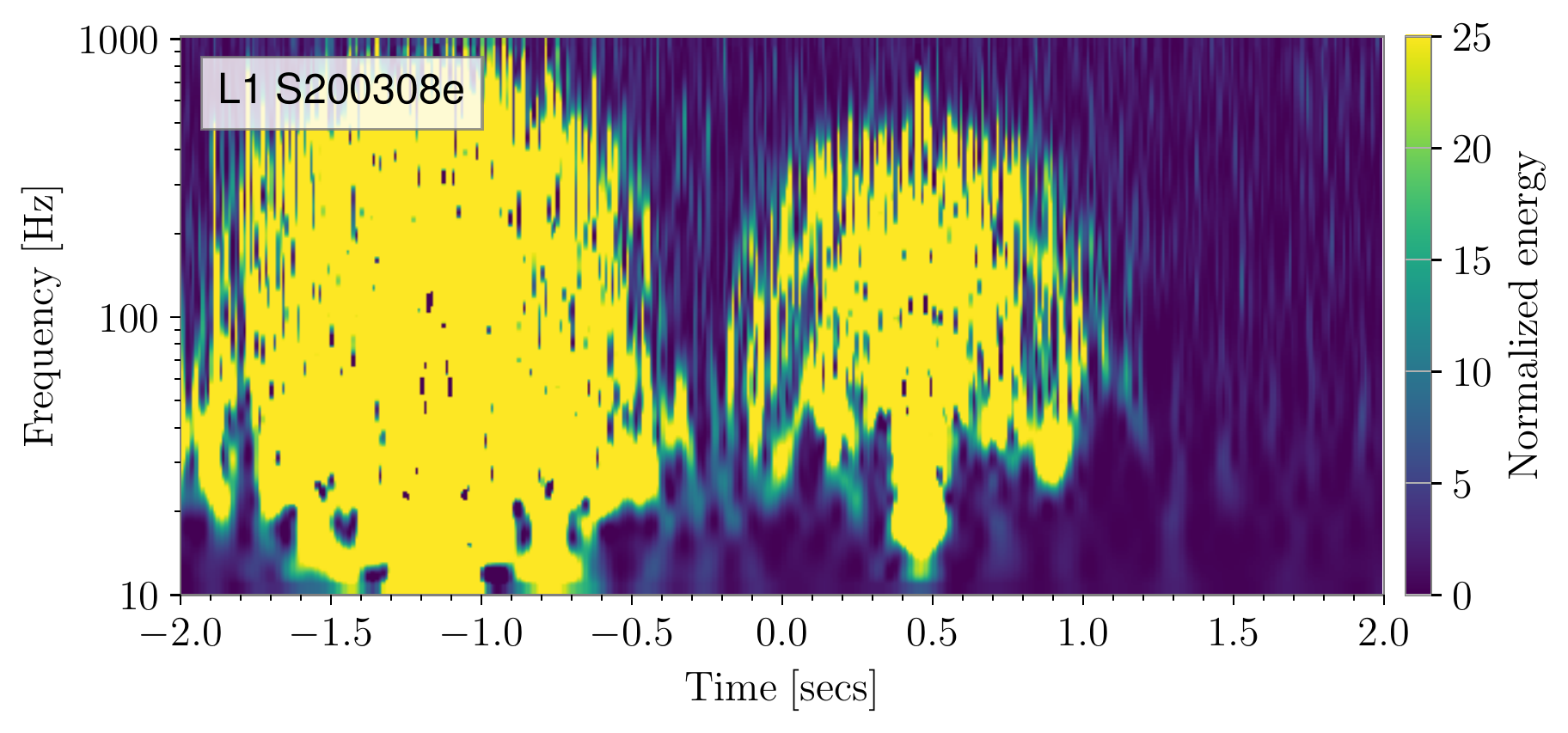}
    \caption{S200308ae}
    \label{fig:third}
\end{subfigure}
\hfill
\begin{subfigure}{0.45\textwidth}
    \includegraphics[width=\textwidth]{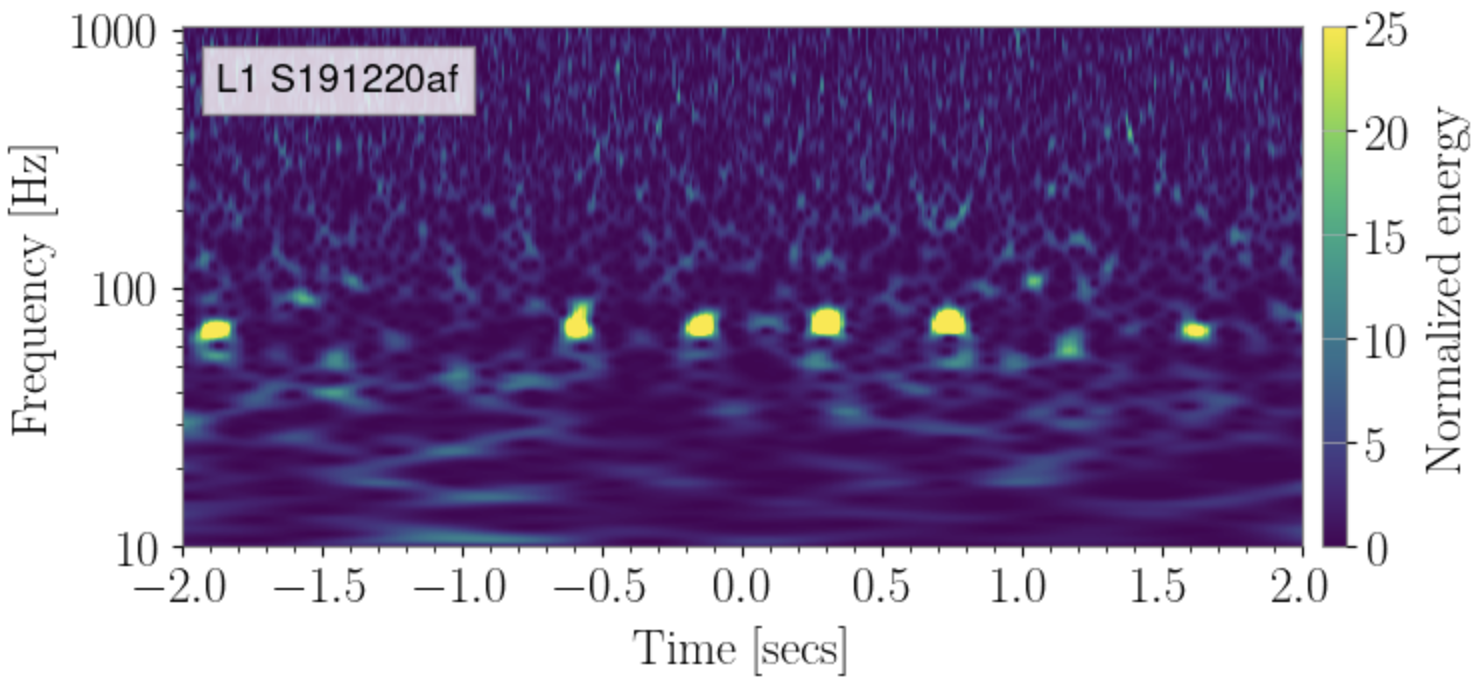}
    \caption{S191220af}
    \label{fig:second}
\end{subfigure}
        
\caption{O3 retracted events flagged as ``Noise" by QoQ test. The last event S191220af is classfied as ``Signal + Noise".}
\label{fig:figures}
\end{figure}

\begin{table}[h]
\centering
    \begin{tabular}{|c|c|c|c|c|c|c|c|c|} 
\hline
 \multicolumn{1}{c}{IFO} & \multicolumn{1}{c}{Freq band [Hz]} & \multicolumn{1}{c}{t1} & \multicolumn{1}{c}{t2} & \multicolumn{1}{c}{t3} & \multicolumn{1}{c}{t4} & \multicolumn{1}{c}{t5} &\multicolumn{1}{c}{QoQ Flag} &\multicolumn{1}{c}{Superevent}\\ 
 \hline
L1 & $10-100 $ & $2.0$ &  $0.7$ & $0.0$	&  $0.0$ &  $0.0$ & Signal $+$ Noise& $S190513bm $ \cite{LIGOScientific:2021djp,GCN24522}\\
\hline
H1 & $10-100 $ &  $0.1$ &	$0.1$ &	$0.$0 &	$0.0$ &	$0.0$  & Signal $+$ Noise & $S191109d$ \cite{LIGOScientific:2021djp,GCN26202}\\
\hline
L1 & $10-100 $  & $0.0$ &  $0.0$ & $0.0$ &  $0.3$ &  $0.0$  & Signal $+$ Noise & $S200224ca$ \cite{LIGOScientific:2021djp,GCN27184}\\
\hline
H1 & $10-100 $ & $0.0$ &  $0.0$ & $0.0$	&  $0.0$ &  $0.4$ & Signal $+$ Noise & $\textbf{S191127p}$ \cite{LIGOScientific:2021djp}\\
\hline 
L1 &  $10-100 $ & $0.0$ &  $0.0$ & $0.0$	&  $0.0$ &  $0.5$  & Signal $+$ Noise & $\textbf{S200216br}$ \cite{LIGOScientific:2021djp}\\
\hline
H1 & $10-100 $  & $0.0$ &  $1.7$ & $0.0$	&  $0.0$ &  $0.0$ & Signal $+$ Noise & $\textbf{S190803e} $ \cite{LIGOScientific:2021djp}\\
\hline
L1 & $10-100 $ & $1.9$ &  $0.4$ & $0.0$	&  $0.0$ &  $0.0$  & Signal $+$ Noise & $\textbf{S190514n}$\cite{LIGOScientific:2021djp}\\
\hline
L1 & $10-100 $ & $8.6$ &  $3.7$ & $0.0$	&  $1.3$ &  $2.5$  &Noise & $S191213g$\cite{LIGOScientific:2021djp,GCN26399}\\
\hline
\end{tabular}

    \caption{QoQ Non Overlapping method analysis of the O3 Catalog events. The table shows the pixel occupancy values of the 7 out of the 74 Catalog events classified as ``Signal + Noise" and 1 out of the 56 O3 online events classified as ``Noise" by the QoQ test. The candidate names in bold were not reported in low latency and were found in offline analyses. None of these 7 events were flagged by the Overlapping method, as shown in Fig \ref{fig:flowchart} and thus are classified as signal with transient noise. 4 of these events (S190513bm, S190514n, S191109d and S191127p) required noise mitigation \cite{LIGOScientific:2021djp, Abbott:2020niy}. The last event in the table, S191213g was classifed as ``Noise" as the L1 data is contaminated by scattering noise just before and after the event time.}
    \label{tab:O3catpixvalsnonovlp} 
\end{table}

\begin{figure}[h]
\centering
\begin{subfigure}{0.45\textwidth}
    \includegraphics[width=\textwidth]{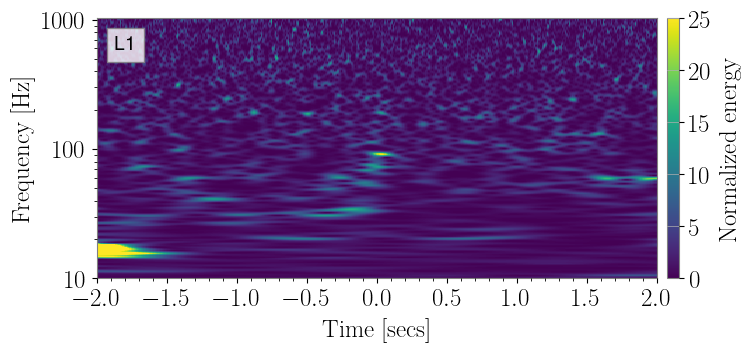}
    \caption{S190513bm}
\end{subfigure}
\hfill
\begin{subfigure}{0.45\textwidth}
    \includegraphics[width=\textwidth]{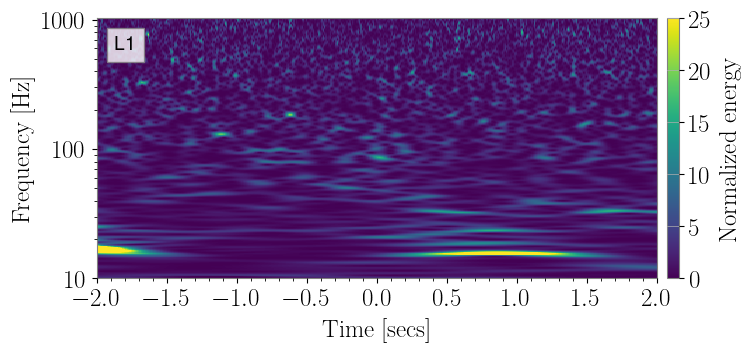}
    \caption{S190514n}
\end{subfigure}

\begin{subfigure}{0.45\textwidth}
    \includegraphics[width=\textwidth]{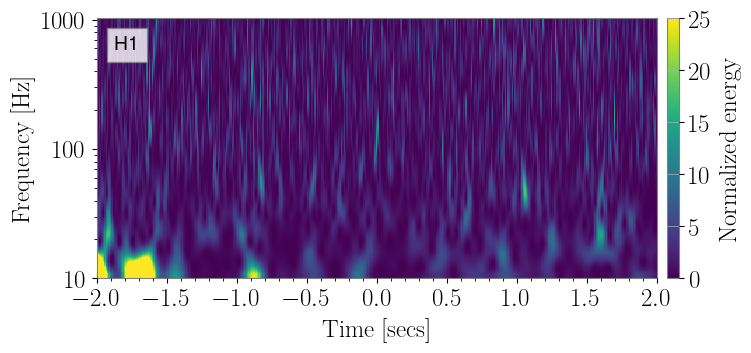}
    \caption{S190803e}
\end{subfigure}
\hfill
\begin{subfigure}{0.45\textwidth}
    \includegraphics[width=\textwidth]{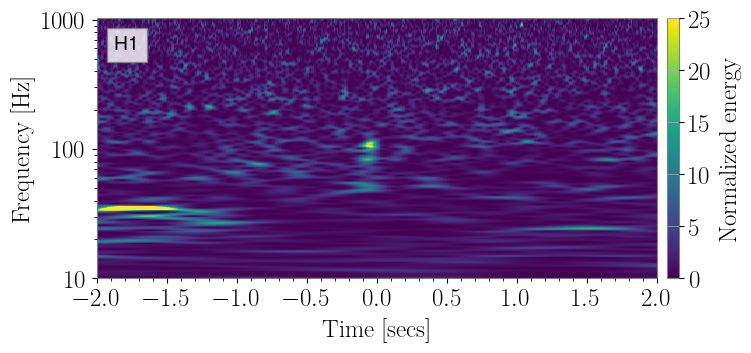}
    \caption{S191109d}
\end{subfigure}

\begin{subfigure}{0.45\textwidth}
    \includegraphics[width=\textwidth]{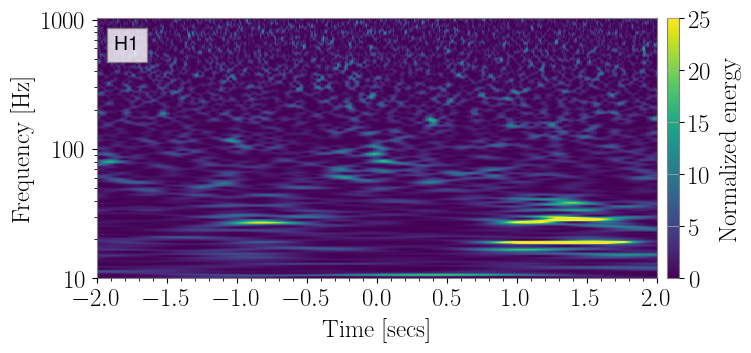}
    \caption{S191127p}
\end{subfigure}
\hfill
\begin{subfigure}{0.45\textwidth}
    \includegraphics[width=\textwidth]{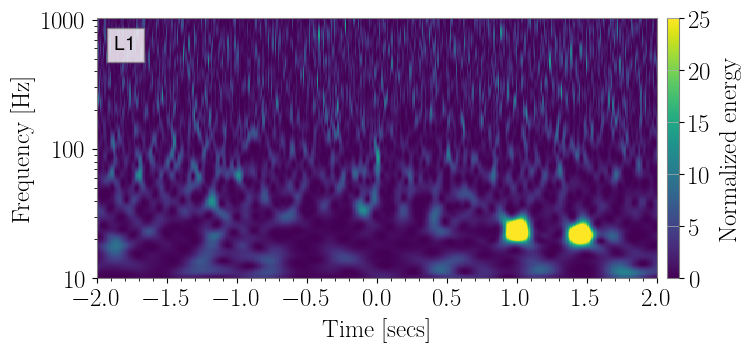}
    \caption{S200216br}
\end{subfigure}

\begin{subfigure}{0.45\textwidth}
    \includegraphics[width=\textwidth]{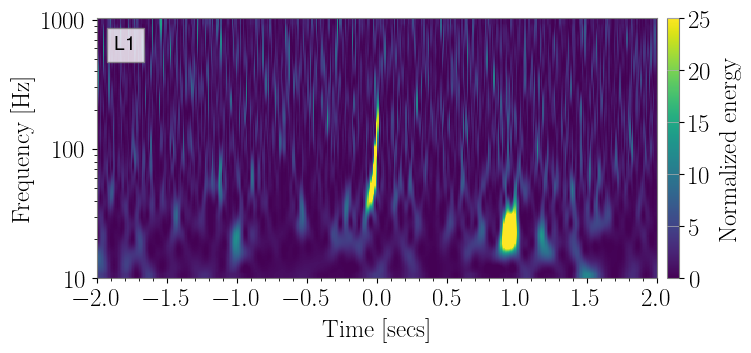}
    \caption{S200224ca}
\end{subfigure}
\hfill
\begin{subfigure}{0.45\textwidth}
    \includegraphics[width=\textwidth]{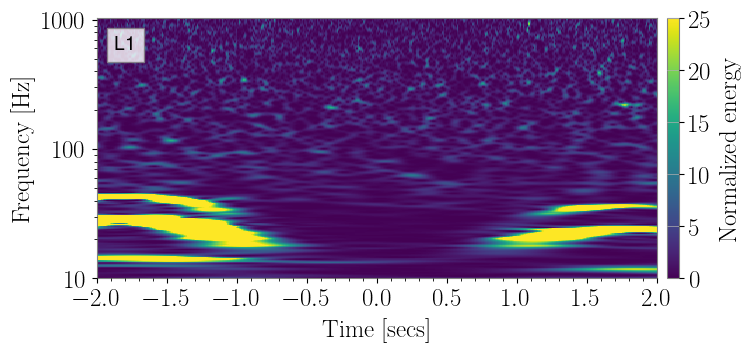}
    \caption{S191213g}
\end{subfigure}

\caption{O3 Catalog events flagged as ``Signal + Noise" and ``Noise" by QoQ test. The first 7 events shown here are classified as ``Signal + Noise" while the event S191213g is classifed as ``Noise'' due the presence of high SNR scattering arches in the data. }
\label{fig:figure_appendixO3}
\end{figure}

\newpage

\bibliographystyle{iopart-num}
\bibliography{qoq.bib}

\end{document}